\documentclass[runningheads]{llncs}
\usepackage[T1]{fontenc}

\usepackage{amsmath}
\usepackage{amssymb}
\usepackage{url}
\usepackage{graphicx}
\usepackage{amscd}
\usepackage{stmaryrd}
\usepackage[usenames,dvipsnames,svgnames,table]{xcolor}
\usepackage{todonotes}
\usepackage{hyperref}
\usepackage{cleveref}
\usepackage{mdframed}
\usepackage{extarrows}
\usepackage{mathtools}
\usepackage{thm-restate}
\usepackage{quiver}

\usepackage{color}

\newenvironment{qroof}{\begin{proof}}{\qed\end{proof}}

\newcommand{\bigslant}[2]{{\raisebox{.2em}{$#1\negthinspace$}\left/\raisebox{-.2em}{$#2$}\right.}}

\newcommand{\Sierp}{\mathbb{S}}

\newcommand{\Cantor}{{\{0,1\}^\mathbb{N}}}

\newcommand{\id}{\textrm{id}}
\newcommand{\dom}{\operatorname{dom}}

\newcommand{\Baire}{{\mathbb{N}^\mathbb{N}}}
\newcommand{\hide}[1]{}
\newcommand{\mto}{\rightrightarrows}

\newcommand{\leqW}{\leq_{\textrm{W}}}
\newcommand{\eqW}{=_{\textrm{W}}}
\newcommand{\leW}{\le_{\textrm{W}}}

\newcommand{\pow}{\mathcal{P}}

\newcommand{\partto}{\rightharpoonup}

\newcommand{\interp}[1]{\llbracket #1 \rrbracket}
\newcommand{\slice}[2]{\bigslant{#1}{#2}}

\newcommand\cL{\mathcal{L}}
\newcommand\cS{\mathcal{S}}

\newcommand{\Hom}[3]{\left[#2,#3\right]_{#1}}
\newcommand\Container{\mathsf{Cont}}

\newcommand\epito{\twoheadrightarrow}
\newcommand{\Asm}{\mathsf{Asm}}
\newcommand{\Mod}{\mathsf{Mod}}
\newcommand{\pAsm}{\mathsf{pAsm}}
\newcommand{\pMod}{\mathsf{pMod}}
\newcommand{\code}[1]{\lceil #1 \rceil}
\newcommand{\Set}{\mathsf{Set}}
\newcommand{\Cat}{\mathsf{Cat}}
\newcommand{\realizes}{\Vdash}
\newcommand{\KVPCA}{\mathcal{K}_2}
\newcommand{\KVPCAR}{\mathcal{K}_2^{\mathrm{rec}}}

\newcommand{\posetOf}[1]{\bigslant{#1}{\leftrightarrow}}
\newcommand{\carrier}[1]{|#1|} %
\newcommand\suppewp[1]{{\|#1\|}}
\newcommand\tuple[1]{\left\langle #1 \right\rangle}
\newcommand\cotuple[1]{{\left[ #1 \right]}}

\newcommand\longto{\longrightarrow}
\newcommand\tensor\otimes

\newcommand\fiberwiseprojective{fibrewise projective}
\newcommand\Fiberwiseprojective{Fibrewise projective}
\newcommand\fpContainer{\Container_{\mathsf{fp}}}
\newcommand\incopr{\mathsf{in}}
\newcommand\coproj\incopr
\newcommand\cod{\mathrm{cod}}
\newcommand\op{\mathrm{op}}

\newcommand\monoto\rightarrowtail

\newcommand{\ltW}{<_\mathrm{W}}

\ifdefined\cA
\else
\newcommand\cA{\mathcal{A}}
\fi
\ifdefined\cB
\else
\newcommand\cB{\mathcal{B}}
\fi
\ifdefined\cC
\else
\newcommand\cC{\mathcal{C}}
\fi
\ifdefined\cD
\else
\newcommand\cD{\mathcal{D}}
\fi
\ifdefined\cE
\else
\newcommand\cE{\mathcal{E}}
\fi
\ifdefined\cF
\else
\newcommand\cF{\mathcal{F}}
\fi
\ifdefined\cG
\else
\newcommand\cG{\mathcal{G}}
\fi
\ifdefined\cH
\else
\newcommand\cH{\mathcal{H}}
\fi
\ifdefined\cI
\else
\newcommand\cI{\mathcal{I}}
\fi
\ifdefined\cJ
\else
\newcommand\cJ{\mathcal{J}}
\fi
\ifdefined\cK
\else
\newcommand\cK{\mathcal{K}}
\fi
\ifdefined\cL
\else
\newcommand\cL{\mathcal{L}}
\fi
\ifdefined\cM
\else
\newcommand\cM{\mathcal{M}}
\fi
\ifdefined\cN
\else
\newcommand\cN{\mathcal{N}}
\fi
\ifdefined\cO
\else
\newcommand\cO{\mathcal{O}}
\fi
\ifdefined\cP
\else
\newcommand\cP{\mathcal{P}}
\fi
\ifdefined\cQ
\else
\newcommand\cQ{\mathcal{Q}}
\fi
\ifdefined\cR
\else
\newcommand\cR{\mathcal{R}}
\fi
\ifdefined\cS
\else
\newcommand\cS{\mathcal{S}}
\fi
\ifdefined\cT
\else
\newcommand\cT{\mathcal{T}}
\fi
\ifdefined\cU
\else
\newcommand\cU{\mathcal{U}}
\fi
\ifdefined\cV
\else
\newcommand\cV{\mathcal{V}}
\fi
\ifdefined\cW
\else
\newcommand\cW{\mathcal{W}}
\fi
\ifdefined\cX
\else
\newcommand\cX{\mathcal{X}}
\fi
\ifdefined\cY
\else
\newcommand\cY{\mathcal{Y}}
\fi
\ifdefined\cZ
\else
\newcommand\cZ{\mathcal{Z}}
\fi
\ifdefined\bA
\else
\newcommand\bA{\mathbb{A}}
\fi
\ifdefined\bB
\else
\newcommand\bB{\mathbb{B}}
\fi
\ifdefined\bC
\else
\newcommand\bC{\mathbb{C}}
\fi
\ifdefined\bD
\else
\newcommand\bD{\mathbb{D}}
\fi
\ifdefined\bE
\else
\newcommand\bE{\mathbb{E}}
\fi
\ifdefined\bF
\else
\newcommand\bF{\mathbb{F}}
\fi
\ifdefined\bG
\else
\newcommand\bG{\mathbb{G}}
\fi
\ifdefined\bH
\else
\newcommand\bH{\mathbb{H}}
\fi
\ifdefined\bI
\else
\newcommand\bI{\mathbb{I}}
\fi
\ifdefined\bJ
\else
\newcommand\bJ{\mathbb{J}}
\fi
\ifdefined\bK
\else
\newcommand\bK{\mathbb{K}}
\fi
\ifdefined\bL
\else
\newcommand\bL{\mathbb{L}}
\fi
\ifdefined\bM
\else
\newcommand\bM{\mathbb{M}}
\fi
\ifdefined\bN
\else
\newcommand\bN{\mathbb{N}}
\fi
\ifdefined\bO
\else
\newcommand\bO{\mathbb{O}}
\fi
\ifdefined\bP
\else
\newcommand\bP{\mathbb{P}}
\fi
\ifdefined\bQ
\else
\newcommand\bQ{\mathbb{Q}}
\fi
\ifdefined\bR
\else
\newcommand\bR{\mathbb{R}}
\fi
\ifdefined\bS
\else
\newcommand\bS{\mathbb{S}}
\fi
\ifdefined\bT
\else
\newcommand\bT{\mathbb{T}}
\fi
\ifdefined\bU
\else
\newcommand\bU{\mathbb{U}}
\fi
\ifdefined\bV
\else
\newcommand\bV{\mathbb{V}}
\fi
\ifdefined\bW
\else
\newcommand\bW{\mathbb{W}}
\fi
\ifdefined\bX
\else
\newcommand\bX{\mathbb{X}}
\fi
\ifdefined\bY
\else
\newcommand\bY{\mathbb{Y}}
\fi
\ifdefined\bZ
\else
\newcommand\bZ{\mathbb{Z}}
\fi

\begin{document}
\title{Weihrauch problems as containers}

\author{Cécilia Pradic\orcidID{0000-0002-1600-8846} \and
Ian Price\orcidID{0009-0009-4112-3385}}
\authorrunning{C. Pradic \& I. Price}
\institute{
Swansea University
}

\maketitle
\begin{abstract}
We note that Weihrauch problems can be regarded as containers over the category
of projective represented spaces and that Weihrauch reductions correspond
exactly to container morphisms. We also show that Bauer's extended Weihrauch
degrees and the posetal reflection of containers over partition assemblies are
equivalent.
Using this characterization, we show how a number of operators over Weihrauch
degrees, such as the composition product, also arise naturally from the
abstract theory of polynomial functors.
\end{abstract}

\section{Introduction}

Weihrauch reducibility allows to compare the computational
strength of partial multi-valued functions over Baire space, which are thus
called \emph{Weihrauch problems}. Much like Reverse Mathematics, this can be
used as a framework to gauge the relative power of certain $\Pi^1_2$ statements
such as ``for every sequence in $\Cantor$, we can produce a cluster
point''~\cite{BGMBW12}. 

Concretely, a partial multi-valued function $f :\subseteq X \mto Y$ is a map
$f : X \to \pow(Y)$. Its domain $\dom(f) \subseteq X$ is the set
of elements such that $f(x) \neq \emptyset$.
When $X = Y = \Baire$, we call such $f$ \emph{Weihrauch problems} and
define reduction between those as follows.

\begin{definition}[{\cite{lmpsv}\footnote{Note that the official definition
in other papers in the literature may differ in two ways: problems may allowed to range
over arbitrary represented spaces and the definition of ``being a reduction''
might involve quantifying over sections of the target multi-valued function.
The latter aspect does not matter in presence of $2^{\aleph_0}$-choice (otherwise it
captures less reductions than the other definition)~\cite[Proposition 3.2]{BGPsurvey}.
For the former, the usage
of arbitrary represented spaces comes with the allowance for $(\varphi, \psi)$
to be multi-valued, which means a problem $P$ is equivalent to its version
defined on codes (which are assumed to be elements of Baire space).}}]
\label{def:weihrauchReducibility}
If $f$ and $g$ are two Weihrauch problems,
$f$ is said to be \emph{Weihrauch reducible to} $g$ if there exists
a pair of partial type 2 computable maps $(\varphi, \psi)$ such that
$\varphi$ is a map $\dom(f) \to \dom(g)$ and for every $i \in \dom(f)$
and $j \in g(\varphi(i))$, $\psi(i,j)$ is defined and belongs to $f(i)$.
\end{definition}

\begin{figure}
\begin{center}
  \includegraphics[scale=0.5]{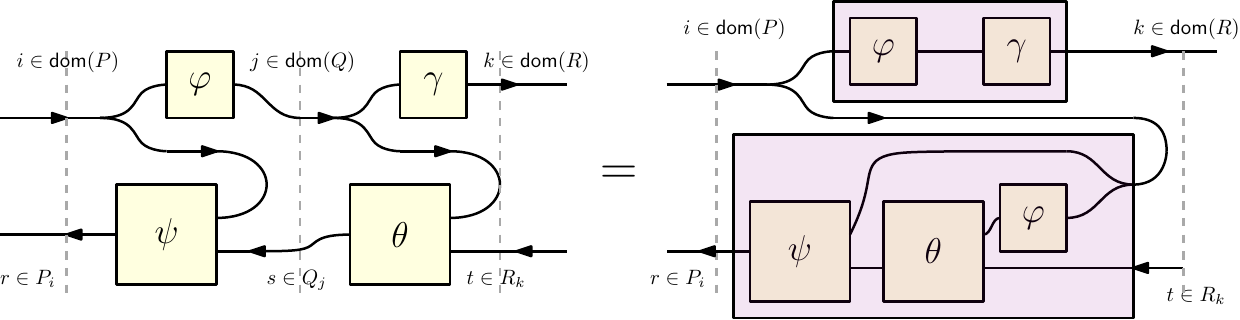}
\end{center}

\caption{String diagram representing the composition of Weihrauch reductions from $P$ to $Q$ and $Q$ to $R$.}
\label{fig:redcompo}

\end{figure}

We write $f \leW g$ when a reduction from $f$ to $g$ exists.
Since Weihrauch reductions compose as depicted in~\Cref{fig:redcompo},
Weihrauch problems and reductions form a preorder. Equivalence classes of
problems are called Weihrauch degrees.

\begin{example}
We can encode the following as Weihrauch problems:
\begin{itemize}
\item $\mathsf{LPO}$
(``given a bit sequence, tell me if it has a $1$''), defined by
\[ \mathsf{LPO}(p) = \{ 0^\omega \mid p = 0^\omega \} \cup \{ 1^\omega \mid p \in \Cantor, p \neq 0^\omega \} \]
\item $\mathsf{LPO}'$ (``given a bit sequence, tell me if it has infinitely many $1$s''), defined by
\[ \mathsf{LPO}'(p) =
  \{ 0^\omega \mid p \in \Cantor, \forall^\infty n. \; p_n = 0 \} \cup \{ 1^\omega \mid p \in \Cantor, \exists^\infty n. \; p_n = 1 \} \]
\item $\mathsf{KL}$: ``given an infinite finitely branching tree $t$, give me an infinite path through $t$'' can also be encoded
  as a Weihrauch problem, modulo a standard embedding $2^{\mathbb{N}^{<\omega}} \hookrightarrow \Baire$.
\item $\mathsf{WKL}$ is the restriction of $\mathsf{KL}$ to trees containing only binary words.
\end{itemize}
We have $\mathsf{LPO} \ltW \mathsf{LPO}'$, $\mathsf{LPO} \ltW \mathsf{KL}$ and $\mathsf{WKL} \ltW \mathsf{KL}$.
  $\mathsf{KL}$ and $\mathsf{LPO'}$ are incomparable, as well as $\mathsf{LPO}$ and $\mathsf{WKL}$.
\end{example}

The Weihrauch degrees actually form a distributive lattice, where meets and
joins can be regarded as natural operators on problems: the join $P \sqcup Q$
allows to ask a question either to $P$ or $Q$ and get the relevant answer,
while $P \sqcap Q$ requires to ask one question to each and get only one answer.
Many other natural operators have been introduced on the Weihrauch lattice,
including a parallel product $P \times Q$ (ask questions to both $P$ and $Q$,
get both answers), a composition operator $P \star Q$ (ask a question to $Q$ and then, depending on the answer
you got, a question to $P$) and many others, including residuals and fixpoints
of other operators~\cite[Definition 1.2]{BGPsurvey}.

The definition of a number of these operators often does not seem to depend much on
the specifics of type 2 computability. They also tend to satisfy nice equational
theories whose soundness can solely be established through abstract nonsense.
It has also been remarked that Weihrauch reducibility shares a likeness with
G\"odel's Dialectica interpretation~\cite[resp. \S 1.9 and 6]{BGPsurvey,TVdP22}.
This hints that ``the
category of Weihrauch reductions'' should be
constructible by applying a functorial construction $\Cat \to \Cat$ (where $\Cat$ is the category of all locally small
categories) to a somewhat canonical category of spaces and type 2 computable
functions between them.

In this paper, we observe that this construction
has been known as the category of \emph{containers}~\cite{containers03}\footnote{
While the name container comes from research on generic programming,
they sometimes also are known under different names in other communities.
They may be called \emph{polynomials} due to their correspondence
with polynomial functors over locally cartesian closed categories~\cite{GKpoly}.
} over projective represented spaces (that is, the full
subcategory of represented spaces containing all subspaces of $\Baire$).

\begin{restatable}{theorem}{mainthmwei}
\label{thm:mainthmwei}
The Weihrauch degrees are isomorphic to the posetal reflection of the category of
answerable\footnote{Meaning those that can be represented by
  surjections between carriers, or equivalently, those pullback-stable epimorphisms; see~\Cref{def:answerable}.}
containers over the category of projective represented spaces and (type 2) computable functions.
\end{restatable}

The proof consists mostly of unpacking definitions; let us
try to give a rough idea of its content.
Containers over a category $\cC$ with pullbacks are bundles, that is, morphisms
in $\cC$. A morphism from the container $P : X \to U$ to $Q : Y \to V$ consists of a forward map
from the base $U$ of $P$ to the base $V$ of $Q$ and a backward map
$Y \times_V U \to X$ involving the total spaces $X$ and $Y$.
One can turn multi-valued functions $f$ into
the bundle $F : \{ (x, y) \mid y \in f(x) \} \xrightarrow{\pi_1} \dom(f)$ to
get half of the correspondence we want and check that this extends to a functor:
Weihrauch reductions \emph{are} morphisms between the containers built this way.

\begin{example}
One can check that the container corresponding to $\mathsf{LPO}$
is
\[ \{(0^\omega, 0^\omega)\} \cup
  \{(p, 1^\omega) \mid p \in \Cantor \setminus \{0^\omega\}\} \longrightarrow
  \Cantor \]
It is equivalent (in the sense that there are morphisms both ways) to
the surjective bundle $l : \bN \epito \bN_\infty \subseteq 2^\bN$ defined by $l(0) = 0^\omega$ and $l(n+1) = 0^n10^\omega$.
\end{example}

This correspondence applies not only to Weihrauch reducibility,
but also to some closely connected variants. One is continuous Weihrauch
reducibility, the notion obtained by replacing ``computable'' by
``continuous'' in~\Cref{def:weihrauchReducibility}, for which the theorem
can be adapted straightforwardly as follows.

\begin{restatable}{theorem}{mainthmweic0}
\label{thm:mainthmweic0}
The continuous Weihrauch degrees are isomorphic to the posetal reflection of the category of answerable
containers over projective represented spaces and continuous functions.
\end{restatable}

Another closely related notion are the extended Weihrauch degrees introduced
in~\cite{Bauer22} by means of a more general notion of reduction between predicates
in a topos. Again a correspondence can be made with containers over a
relatively natural category of projectives,
although this requires a bit of massaging.

\begin{restatable}{theorem}{mainthmweiext}
\label{thm:mainthmweiext}
Assuming the axiom of choice, the extended Weihrauch degrees over a filtered PCA $(\bA', \bA)$ are isomorphic to the posetal reflection of the category of containers over partitioned assemblies $\pAsm(\bA', \bA)$.
\end{restatable}

In making those folklore correspondences explicit, we hope among other things to link
the literature on Weihrauch reducibility and containers. After providing
the necessary background definitions (\Cref{sec:background}) and proving the
theorems stated thus far (\Cref{sec:polynomials}), we discuss some of the
algebraic operators of the Weihrauch lattice in the context of the theory of
containers (\Cref{sec:structure}). We will only focus on the two lattice
operations, the parallel product and the composition product. While we will
obtain clean correspondences to known functorial constructions for the first
three, we will hit a snag when it comes to the composition product. This is
due to the fact that the categories of spaces we are taking containers
over are only \emph{weakly} locally cartesian closed.
This means that the composition product will only be a quasi-functor, which
complexifies its systematic study; we leave that for further work.

\paragraph*{Related work}

The idea of regarding a bundle as a problem to be solved by finding one of its
section is old one that predates the ``container'' terminology. For instance,
Hirsch~\cite[Definition 3.4]{HirschThesis90} defines an equivalent category
to study the topological complexity of problems and reductions between those.
This perspective also already appeared in the
literature on Weihrauch problems (see for instance~\cite[Remark 2.8]{Kihara16BP}),
although most of the recent efforts we are aware of to ``categorify'' Weihrauch
reducibility and operators on problems tend to use other tools instead.

One such natural categorical construction that
captured Weihrauch degrees  is the restriction
of Bauer's extended Weihrauch
problems~\cite{Bauer22} to objects that are actually Weihrauch problems, which
are characterized as the $\neg\neg$-dense predicates over modest sets.
Interestingly, Ahman and Bauer also linked generalized Weihrauch reducibility
to containers~\cite{AhmanBauer24}, but by way of the more general notion of
instance reducibility that works over families of truth values. They get back
to generalized Weihrauch predicated by specializing their constructions to
assemblies which do have a natural truth value object which is not partitioned
while here we work directly with bundles of partitioned assemblies.

Trotta et. al.~also formally linked (extended) Weihrauch reducibility with
the Dialectica interpretation~\cite{TVdP22}, which can be regarded
bicompletions of fibrations by simple products and sums~\cite{hofstra11mlhc}.
Aside from the fact that they work in a posetal setting throughout,
it is interesting to note that the category of containers over $\cC$ can be
recovered by completing the terminal fibration over $\cC$ by arbitrary
products and then sums and taking the fiber over $1$.

Pauly also studied a generic notion of reducibility that encompasses Weihrauch
reducibility, starting from categories of multivalued
functions~\cite{PaulyManyOneRedAbstract}, in which he derived the lattice
operators as well as finite parallelizations in a generic way. In contrast,
we try to stay as far away as possible from multi-valued functions as a notion
of morphism in this paper.

\section{Categories for computable analysis}
\label{sec:background}

We review briefly some standard terminology from category theory and realizability
that we will use in this paper. In the rest of this paper we will favor
terminology from realizability rather than computable analysis for convenience
as it allows us to state~\Cref{thm:mainthmweiext}.
Aside from the use of partial combinatory algebras, the main terminological
differences are summarized in \Cref{fig:catNames}. We assume that
the reader is familiar with the notion of what are categories, functors,
natural transformations, finite (co)limits and what is type-2 computability.

\subsection{Categories for type-2 computability}

Since we are going to be borrowing notions from realizability, we are going
to use a definition based on the notion of partial combinatory algebras (PCAs).
Without going into details (which the interested reader can find in e.g.~\cite{vanOosten}),
a PCA is given by a set $\bA$ (its \emph{carrier}) and a partial binary operation ${\cdot} : \bA^2 \partto \bA$.
The intuition is that
elements of $\bA$ are to be regarded as codes for programs and that $\cdot$
denotes function application; as such we simply call it application in the
sequel.
In addition to having this operation, a PCA $\bA$ is required to have certain distinguished
elements $\mathsf{s}$ and $\mathsf{k}$ satisfying certain equations, essentially
so that the untyped $\lambda$-calculus is interpretable in $\bA$.

\begin{example}
The PCA $\mathcal{K}_1$ (Kleene's first algebra) has carrier $\bN$ and the
application $n \cdot m$ is defined to be the output of running the $n$th 
Turing machine on $m$, otherwise it is undefined.
\end{example}

One of the PCAs we will be concerned with will be $\mathcal{K}_2$ (Kleene's second
algebra). The carrier of $\mathcal{K}_2$ is $\Baire$, and the intuition is
that we can regard elements of Baire space as partial functions $\Baire \partto \bN$ as
follows: up to squinting, we can regard elements of $\Baire$ as countably
branching trees whose leaves are labelled by natural numbers. One can
then define the partial function represented by such a tree $t$ as mapping
$p \in \bN$ to $n \in \bN$ if and only if following the path $p$ in $t$
leads to a leaf labelled by $n$ (if $p$ is an infinite branch of $t$, the
function is undefined). The obtained partial function is continuous and
it can be shown that all partial continuous functions can be defined this way.
The application of the PCA, which is meant to
regard elements of $t$ as maps $\Baire \partto \Baire$ can be defined on this
basis using $\Baire \cong \Baire \times \bN$.

For the purpose of computable analysis, we need to use computable continuous
functions in a setting where all elements of $\Baire$ are in the background.
For this we need the notion of a \emph{filtered PCA}
$(\bA', \bA)$~\cite[Definition 1.5]{vanoosten2016classicalrelativerealizability},
which is a pair where $\bA$ is a PCA and $\bA'$ is a subPCA of $\bA$ (i.e., it
is stable by $\cdot$ and contains the distinguished $\mathsf{k}$ and $\mathsf{s}$,
i.e., the interpretation of untyped $\lambda$-calculus with parameters in $\bA'$
using the structure of $\bA$ lands in $\bA'$).

The one filtered PCA we shall use in this paper is $(\KVPCAR, \KVPCA)$ where
$\KVPCAR$ consists of the set of computable elements of $\Baire$. It can
be checked that they correspond, through applications, to functions computable
by type-2 Turing machines.
When defining a category $\cC(\bA',\bA)$ parameterized by a filtered PCA
$(\bA', \bA)$, we may write $\cC(\bA)$ for $\cC(\bA,\bA)$ for brevity's sake.

\begin{figure}
\begin{center}
\begin{tabular}{|l||c|c|c|}
\hline
Realizability &
\begin{tabular}{c}
  modest sets\\
  ($\Mod$)
\end{tabular} &
\begin{tabular}{c}
assemblies\\
($\Asm$)
\end{tabular}
&
\begin{tabular}{c}
partitioned modest sets\\
($\pMod$)
\end{tabular}
\\
\hline

  \begin{tabular}{l}
Computable \\analysis
\end{tabular} & \begin{tabular}{c}
represented \\ spaces
\end{tabular}
&
\begin{tabular}{c}
multi-represented\\
spaces
\end{tabular}
&
\begin{tabular}{c}
  represented subspaces of $\Baire$\\
  (projective represented spaces)
\end{tabular}
\\
\hline
\end{tabular}
\end{center}
\caption{Categories used in realizability and their counterpart, once
  they are specialized to the filtered PCA $(\KVPCAR,\KVPCA)$,
in computable analysis.}
\label{fig:catNames}
\end{figure}

With this machinery in place, we can start to define categories of sets
represented by codes in PCAs. We begin with the more general notion we will
encounter in this paper and decline the useful subcategories we will discuss.

\begin{definition}
An assembly over a PCA $\bA$
is a pair $(X, \realizes_X)$
where $X$ is a set called the \emph{carrier} and $\realizes_X \subseteq \bA \times X$ is a relation
whose image is $X$.

A map $f$ between the carriers of the assemblies $(X, \realizes_X)$
and $(Y, \realizes_Y)$ is \emph{tracked} by a code $\code{f} \in \bA$ if,
whenever $e \realizes_X x$, we have $\code{f} \cdot e \realizes_Y f(x)$.

Assuming $(\bA', \bA)$ is a filtered PCA, the category $\Asm(\bA', \bA)$
has assemblies over $\bA$ as objects and maps tracked by codes in $\bA'$ as
morphisms.
\end{definition}

Given an assembly $A$, write $\carrier{A}$ for its carrier.

\begin{remark}
An object of $\Asm(\KVPCAR,\KVPCA)$ is essentially the same thing as a
multi-represented space~\cite[Remark 5.2]{iljazovic2021computability} and
the morphisms of $\Asm(\KVPCAR,\KVPCA)$ (respectively $\Asm(\KVPCA)$) are the computable maps between them.
\end{remark}

\begin{definition}
Given a filtered PCA $(\bA', \bA)$, the category of modest sets
$\Mod(\bA',\bA)$ is the full subcategory of assemblies $(X, \realizes_X)$
such that any element of $\bA$ realizes at most one element (i.e.,
$e \realizes_X x$ and $e \realizes_X x'$ imply $x = x'$).
\end{definition}

\begin{remark}
The objects of $\Mod(\KVPCA)$ are essentially \emph{represented spaces}.
\end{remark}

\begin{definition}
Call an assembly \emph{partitioned} when each of its elements have at most
  one realizer. We write $\pMod(\bA',\bA)$ and $\pAsm(\bA',\bA)$ for the respective full subcategories of partitioned
  modest sets $\Mod(\bA',\bA)$ and assemblies $\Asm(\bA',\bA)$.
\end{definition}

\begin{remark}
The objects of $\pMod(\KVPCA)$ are essentially \emph{represented subspaces} of $\Baire$,
  which can also be characterized as the regular projective objects of $\Mod(\KVPCA)$ (\Cref{lem:pmod-pass-projectives}).
\end{remark}

\subsection{Notations for category-theoretical notions}
\label{subsec:ctbasics}

We assume familiarity with the notion of cartesian products and coproducts.
We write $\tuple{f_1, f_2} : Z \to A_1 \times A_2$ for the pairing of $f_i : Z \to A_i$
($i \in \{1,2\}$)
and $g : Z \to A_2$ and $\pi_i : A_1 \times A_2 \to A_i$ ($i \in \{1,2\}$.
Dually we write $\cotuple{f_1, f_2} : A_1 + A_2 \to Z$
and $\incopr_i : A_i \to A_1 + A_2$ for the copairing and coprojections.

We will heavily use the notion of pullbacks throughout this paper as well
as standard notation that involves them. A pullback square is a commuting
square as found in the diagram below satisfying the following universal 
property: if there are morphisms $\alpha$, $\beta$ as depicted such that
$f \circ \beta = g \circ \alpha$, then there is a unique $\gamma$
such that $\alpha = k \circ \gamma$ and $\beta = h \circ \gamma$.

\[\begin{tikzcd}
	Z \\
	& {A \times_C B} & B \\
	& A & C
	\arrow["{\exists! \gamma}"', dashed, from=1-1, to=2-2]
	\arrow["\alpha", curve={height=-12pt}, from=1-1, to=2-3]
	\arrow["\beta"', curve={height=12pt}, from=1-1, to=3-2]
	\arrow["k"', from=2-2, to=2-3]
	\arrow["h", from=2-2, to=3-2]
	\arrow["\lrcorner"{anchor=center, pos=0.125}, draw=none, from=2-2, to=3-3]
	\arrow["g", from=2-3, to=3-3]
	\arrow["f"', from=3-2, to=3-3]
\end{tikzcd}\]

In this diagram $A \times_C B$ (together with the projections $h$ and $k$)
is a pullback of $f$ and $g$. We use the notation around the top-left corner of
the commuting square to indicate we intend this to be a pullback in a diagram.
Pullbacks are only determined up to unique isomorphisms.
Concretely, pullbacks in all categories we are interested in here can be built by
taking the cartesian products of the domains of the maps $f$ and $g$ and
restricting to the sets of pairs $(a,b)$ with $f(a) = g(b)$.
When we fix a choice
of pullbacks for each pair of morphisms, we write $f^*(g)$ for the morphism
corresponding to $h$ (read ``the pullback of $g$ along $f$''); $f^*$ actually
extends to a functor $\slice{\cC}{C} \to \slice{\cC}{A}$ where $\slice{\cC}{A}$
(``the slice over $A$'') is the category whose objects are morphisms of $\cC$
with codomain $A$ and morphisms are maps between domains of objects that make
the obvious triangle diagram commute.

We recall that an \emph{epimorphism} is a morphism $f : A \to B$ such that,
whenever we have $h, k : B \to Z$ with $h \circ f = k \circ f$, we actually
know that $h = k$.

\begin{definition}
An epimorphism $e : X \epito I$ is called \emph{pullback-stable} if for
every pullback square

\[\begin{tikzcd}
	\cdot & X \\
	\cdot & I
	\arrow[from=1-1, to=1-2]
	\arrow["{e'}"', from=1-1, to=2-1]
	\arrow["\lrcorner"{anchor=center, pos=0.125}, draw=none, from=1-1, to=2-2]
	\arrow["e", two heads, from=1-2, to=2-2]
	\arrow[from=2-1, to=2-2]
\end{tikzcd}\]

$e'$ is also an epimorphism.
\end{definition}

In topological spaces (and a fortiori in represented spaces), the pullback-stable
epimorphisms can be characterized as the maps that are surjective in $\Set$.
This does not cover all epimorphisms.

\begin{example}
$\{ p \in \Cantor \mid \forall^\infty n. \; p_n = 0\} \subseteq \Cantor$ gives rise to an epimorphism in $\pMod(\KVPCAR,\KVPCA)$,
but that epimorphism is not pullback-stable.
\end{example}

Finally, let us recall that preorders (and a fortiori posets) can be regarded as
categories whose objects are elements of the preorder and with a unique morphism
between two objects $x$ and $y$ if and only if $x \le y$.
Posets form a reflective subcategory of the category of small categories $\Cat$,
which means that the inclusion of posets into categories has a left adjoint
functor $\cC \mapsto \posetOf{\cC}$.
$\posetOf{\cC}$ is the quotient of $\cC$ where objects $X$ and $Y$ are equivalent
when there exists morphisms $X \to Y$ and $Y \to X$.

\section{Reducibility as container morphisms}
\label{sec:polynomials}

\subsection{A container is an internal family}

At the intuitive level, a container in a category $\cC$ is a
family of objects of $\cC$ indexed by an object of $\cC$. Since objects
of $\cC$ are not necessarily sets, to make this formal, one needs to define what
is a family \emph{internal} to a category. This is done by regarding objects of
$\slice{\cC}{I}$ as $I$-indexed families. The basic idea is that in $\Set$,
given a family $(A_i)_{i \in I}$, we can form the disjoint union
$\sum_{i \in I} A_i$ and encode the family in this formalism by taking the
projection $\sum_{i \in I} A_i ~\to~ I$. Conversely, to show that the categorical
point of view captures something equivalent to the usual notion in $\Set$, we
can perform an inverse operation turning any set-theoretic map $f : X \to I$
into the family $(f^{-1}(i))_{i \in I}$.

A fundamental operation one may perform on families is \emph{reindexing}: given
a map of index sets $f : I \to J$ and a family over $J$, one may produce a
family over $I$.
Then the fundamental operation of reindexing families along a map $f : I \to J$
is given by taking a pullback.
\[
\begin{array}{c !\qquad | !\qquad c}
\text{Families of sets} & \text{Families internal to $\cC$} \\
\hline
\\
  \begin{array}{ccc}
    \text{Families over $J$} &\to& \text{Families over $I$} \\
    \left(B_j\right)_{j \in J} &\mapsto& \left(B_{f(i)}\right)_{i \in I} \\
  \end{array}
 & 
\begin{array}{ccc}
\slice{\cC}{J} &\to& \slice{\cC}{I} \\
\begin{tikzcd}
	B \\
	J
	\arrow["\alpha", from=1-1, to=2-1]
\end{tikzcd}
&\mapsto&
\begin{tikzcd}
	{B'} & \textcolor{rgb,255:red,89;green,89;blue,89}{B} \\
	I & \textcolor{rgb,255:red,89;green,89;blue,89}{J}
	\arrow["\alpha", color={rgb,255:red,89;green,89;blue,89}, from=1-2, to=2-2]
	\arrow[color={rgb,255:red,89;green,89;blue,89}, from=2-1, to=2-2]
	\arrow[color={rgb,255:red,89;green,89;blue,89}, from=1-1, to=1-2]
	\arrow["{f^*\alpha}"', from=1-1, to=2-1]
	\arrow["\lrcorner"{anchor=center, pos=0.125}, draw=none, from=1-1, to=2-2]
\end{tikzcd}
\end{array}
\end{array}
\]

For any choice\footnote{Only unique up to unique isomorphism in $\slice{\cC}{J}$.}
of pullbacks, this yield functors
$f^* : \slice{\cC}{J} \to \slice{\cC}{I}$ which we call
\emph{reindexing along $f$}.
Those functor always have a left adjoint $\Sigma_f : \slice{\cC}{J} \to \slice{\cC}{I}$
given by precomposition that corresponds to \emph{sums along $f$}.
In $\Set$, this corresponds to being able to take set-indexed disjoint unions:
given a family $(A_i)_{i \in I}$, $f$ allows us to consider it as a family of
families over $J$ we may define the family of families, 
$((A_i)_{i \in f^{-1}(j)})_{j \in J}$ and then perform the set-theoretic
disjoint union component-wise.
\[
\begin{array}{c !\qquad | !\qquad c}
\text{Families of sets} & \text{Families internal to $\cC$} \\
\hline
\\
  \begin{array}{ccc}
    \text{Families over $I$} &\to& \text{Families over $J$} \\
    \left(A_i\right)_{i \in I} &\mapsto&
    \left(\sum\limits_{i \in f^{-1}(j)} A_i\right)_{j \in J}\\
  \end{array}
 & 
\begin{array}{ccc}
\slice{\cC}{I} &\to& \slice{\cC}{J} \\
\begin{tikzcd}
	A \\
	I
	\arrow["\alpha", from=1-1, to=2-1]
\end{tikzcd}
&\mapsto&
\begin{tikzcd}
	A \\
	I \\
	J
	\arrow["\alpha", from=1-1, to=2-1]
	\arrow["f", from=2-1, to=3-1]
\end{tikzcd}
\end{array}
\end{array}
\]

The existence of right adjoints $\Pi_f$ to reindexing functors $f^*$ is
however not always guaranteed for every category with pullbacks. It is the case if and only if
all slice categories of $\cC$ are cartesian closed, in which case we call
$\cC$ \emph{locally cartesian closed}. In such a case, $\Pi_f$ corresponds
to \emph{products} and is right adjoint to $f^*$.

\[
\begin{array}{c !\qquad | !\qquad c}
\text{Families of sets} & \text{Families internal to $\cC$} \\
\hline
\\
  \begin{array}{ccc}
    \text{Families over $I$} &\to& \text{Families over $J$} \\
    \left(A_i\right)_{i \in I} &\mapsto&
    \left(\prod\limits_{i \in f^{-1}(j)} A_i\right)_{j \in J}\\
  \end{array}
 & 
\begin{array}{ccc}
\slice{\cC}{I} &\to& \slice{\cC}{J} \\
\alpha
&\mapsto&
  (f \circ \alpha)^f
\end{array}
\end{array}
\]

Having access to the functors $\Sigma_f \dashv f^* \dashv \Pi_f$ means that
we can interpret closed type theoretic-expressions that involve equalities,
sums and products over specified objects of $\cC$ as objects of $\cC$; open
expressions containing variables are interpreted as families whose indexing
objects correspond to the variables.

\subsection{The category of containers: morphisms}

As said previously, the objects of the category $\Container(\cC)$ of containers
over $\cC$ will be morphisms in $\cC$. Following
terminology from~\cite{polybook}, for such a container $P : X \to U$, we
will call $U$ the object of \emph{positions} and $X$ the object of
\emph{directions}.
Now the important thing with containers is the notion of morphism that comes
with them. In sufficiently structured categories, they correspond to strong
natural transformations between functors induced by the containers~\cite[Theorem 2.12]{GKpoly}, but they
also admit a more general low-level characterization that is easily seen to
correspond to notions of reducibility\footnote{And was also independently defined for such purposes, see e.g.~\cite[Definition 3.4]{HirschThesis90}.}.
With sets, assuming we have $Q : Y \to V$, $P_u = P^{-1}(u)$ and
$Q_v = Q^{-1}(v)$, a morphism of container from $P$ to $Q$ is given by a pair
of maps
\[ \varphi : U \to V \qquad \text{and} \qquad \psi \in \prod_{u \in U} (Y_{\varphi(u)} \to X_u) \]
Note that the backward map $\psi$ is here typed using dependent products, which
only exist internally in locally cartesian closed categories. We adopt the more
elementary equivalent definition which works in all categories with pullbacks.

\begin{definition}[{\cite[(4)]{GKpoly}}]
 \label{def:containerMorphismRepr}
A morphism representative from a container $P : X \to U$ to $Q : Y \to V$ is
a pair $(\varphi, \psi)$ making the following diagram commute, the rightmost
square being a pullback\footnote{While we give a suggestive type-theoretic name
for the relevant object, it is determined up to unique isomorphism. This is
why we need to quotient morphism representatives; with chosen pullbacks, we
could have dispensed with this step.}:
\[\begin{tikzcd}
  X && \sum\limits_{u \in U} Y_{\varphi(u)} && Y \\
	U && U && V
	\arrow["P"', from=1-1, to=2-1]
	\arrow["\psi"', from=1-3, to=1-1]
	\arrow[from=1-3, to=1-5]
	\arrow[from=1-3, to=2-3]
	\arrow["\lrcorner"{anchor=center, pos=0.125}, draw=none, from=1-3, to=2-5]
	\arrow["Q", from=1-5, to=2-5]
	\arrow[shift left, no head, from=2-1, to=2-3]
	\arrow[shift right, no head, from=2-1, to=2-3]
	\arrow["\varphi"', from=2-3, to=2-5]
\end{tikzcd}\]
Two morphisms representatives $(\varphi, \psi)$ and $(\varphi', \psi')$
from $P$ to $Q$ are defined to be equivalent if and only if $\varphi = \varphi'$
and $\psi \circ \theta = \psi'$ when $\theta$ is the canonical isomorphism between
the domain of $\psi$ to the domain of $\psi'$ that witnesses that they both 
are pullbacks of $Q$ and $\varphi$. A morphism of containers $P \to Q$
is an equivalence class of morphism representatives.
\end{definition}

Given a morphism (representative) $(\varphi, \psi)$, we will call $\varphi$
the forward map and $\psi$ the backward map. Composition is defined essentially
as suggested in~\Cref{fig:redcompo}; more formally, it is determined as
per the categorical diagram in~\Cref{fig:actualcompo}.
We will call $(\varphi, \psi)$ \emph{horizontal} when $\psi$ is the identity
and \emph{vertical} when $\varphi$ is the identity.

\begin{figure}
\[\begin{tikzcd}
	X & \cdot & \cdot & Z \\
	& \cdot & Y \\
	U & U & V & W
	\arrow["P"', from=1-1, to=3-1]
	\arrow["\chi"', color={rgb,255:red,255;green,51;blue,61}, curve={height=6pt}, from=1-2, to=1-1]
	\arrow[from=1-2, to=1-3]
	\arrow[from=1-2, to=2-2]
	\arrow["\lrcorner"{anchor=center, pos=0.125}, draw=none, from=1-2, to=2-3]
	\arrow[from=1-3, to=1-4]
	\arrow["\beta", from=1-3, to=2-3]
	\arrow["\lrcorner"{anchor=center, pos=0.125}, draw=none, from=1-3, to=3-4]
	\arrow["R", from=1-4, to=3-4]
	\arrow["\psi", from=2-2, to=1-1]
	\arrow[from=2-2, to=2-3]
	\arrow[from=2-2, to=3-2]
	\arrow["\lrcorner"{anchor=center, pos=0.125}, draw=none, from=2-2, to=3-3]
	\arrow["Q", from=2-3, to=3-3]
	\arrow[shift right, no head, from=3-1, to=3-2]
	\arrow[shift left, no head, from=3-1, to=3-2]
	\arrow["\varphi"', from=3-2, to=3-3]
	\arrow["\theta"', color={rgb,255:red,255;green,51;blue,61}, curve={height=18pt}, from=3-2, to=3-4]
	\arrow["\alpha"', from=3-3, to=3-4]
\end{tikzcd}\]
\caption{
We say that $(\theta, \chi) : P \to R$ is a composition of the morphism representatives $(\varphi,\psi) : P \to Q$ and $(\alpha,\beta) : Q \to R$.
Implicit in~\Cref{lem:ContainerCat} is that pairs of equivalent representatives admit the same compositions.}
\label{fig:actualcompo}
\end{figure}

\begin{lemma}
\label{lem:ContainerCat}
Assume $\cC$ is a category with pullbacks. Then we can form a category $\cC$
where the objects are maps in $\cC$, morphisms are given as in~\Cref{def:containerMorphismRepr},
identities over $P : X \to U$ are represented by $(\id_U, \id_X)$ and the 
composition is given as~\Cref{fig:actualcompo}.
\end{lemma}

\begin{qroof}
It can be checked by unpacking the definitions that $\Container(\cC)$ is
exactly the total category of the fibration $\cod^\op$ over $\cC$, which is
obtained by taking the opposite of the codomain
fibration over $\cC$ (which exists because of pullbacks);
see~\cite[\S 1 \& \S 5]{streicherfibrations} for details.
\end{qroof}

\subsection{Weihrauch problems are containers over partitioned modest sets}

We are now almost ready to detail how to prove the main theorem, except for one
thing: some of the containers in $\Container(\pMod(\KVPCAR,\KVPCA))$ may
represent families of subsets of $\Baire$ that may have the empty set as
a member. Weihrauch problems do not allow that, so we need to restrict the
class of objects to rule out those containers.

\begin{definition}
\label{def:answerable}
We call a container \emph{answerable} if the underlying map is a pullback-stable epimorphism. 
\end{definition}

As previously mentioned, in $\Asm$ and a fortiori $\pMod$,
the pullback-stable epimorphisms are surjections over the relevant carriers;
this ensures each fiber is non-empty. It will be sometimes interesting in the
sequel to also consider containers of $\Container(\KVPCAR,\KVPCA)$ which
are not answerable - so when needed we will refer to objects of
(the posetal reflection of) $\Container(\KVPCAR, \KVPCA)$ as \emph{slightly extended Weihrauch problems (degrees)}.

\mainthmwei*
\begin{qroof}
Recall that the category of projective represented spaces is isomorphic to
$\pMod(\KVPCAR, \KVPCA)$.
To each answerable container $P : X \to U$ over $\pMod(\KVPCAR, \KVPCA)$,
we can map the $P$ to the Weihrauch problem $\mathsf{w}(P) \mathrel{:\subseteq} \Baire \mto \Baire$
with $\mathsf{w}(P)(p) = \{ q \in \Baire \mid \exists x \in \carrier{X}. \; p \Vdash_U P(x) \wedge
q \Vdash_X x \}$. For any other $Q : Y \to V$ and a reduction representative
$(\varphi, \psi) : P \to Q$, we can find an equivalent one $(\varphi, \psi')$
where the domain of $\psi'$ is actually the set of pairs $(u, y)$ with
$Q(y) = \varphi(u)$ as it is a pullback. Then $(\varphi, \psi')$ is literally
a Weihrauch reduction from $\mathsf{w}(P)$ to $\mathsf{w}(Q)$.

In the other direction, given a Weihrauch problem $f {:\subseteq} \Baire \mto \Baire$,
we have a map $\mathsf{c}(f) : \sum\limits_{u \in \dom(f)} f(u) \xrightarrow{\pi_1} \dom(f)$;
we can regard its domain and codomain as carriers of partitioned modest sets by taking the
  trivial realizability relations $\tuple{e, e'} \Vdash_{\dom(\mathsf{c}(f))} (e,e')$
and $e \Vdash_{\dom(f)} e$.
By definition, if a reduction $f \leW g$ is represented by a pair of computable
maps $(\varphi, \psi)$, we have $\varphi : \dom(f) \to \dom(g)$ and
$\psi : \sum\limits_{u \in \dom(f)} g(\varphi(u)) \to \bigcup\limits_{u \in \dom(f)} f(u)$.
Since $\psi(u,y)$ is meant to give an answer to $u$, by combining $\psi$ with the
projection $\sum\limits_{u \in \dom(f)} g(\varphi(u)) \to \dom(f)$,
we can form a map $\widetilde{\psi} : \sum\limits_{u \in \dom(f)} g(\varphi(u)) \to \sum\limits_{u \in \dom(f)} f(u)$
so that $(\varphi, \psi)$ is a morphism representative $\mathsf{c}(f) \to \mathsf{c}(g)$.

Now, to conclude is suffices to show that $\mathsf{c}(\mathsf{w}(P)) \cong P$
and $\mathsf{w}(\mathsf{c}(f)) \eqW f$ for all $P$ and $f$.

For the first part, for $P : X \to U$, because $P$ is answerable, we
do also have $\dom(\mathsf{w}(P)) = \{ p \in \Baire \mid \exists u \in \carrier{U}. \; p \Vdash_U u\}$;
this is enough to derive an isomorphism $\theta : \dom(\mathsf{w}(P)) \cong U$ in $\pMod(\KVPCAR,\KVPCA)$
which is tracked by the code for the identity. Then we can easily check that
$\mathsf{c}(\mathsf{w}(P))$ is a pullback of $P$ along $\theta$, hence $(\theta, \id)$
and $(\theta^{-1}, \id)$ represent an isomorphism $\mathsf{c}(\mathsf{w}(P)) \cong P$ as expected.

For the second part we have that $\mathsf{w}(\mathsf{c}(f))(p) = \{ \tuple{p, x} \mid x \in f(p)\}$
for every $p \in \Baire$. So the forward component of both reductions we want
can be taken to be the identity, the backward component of the reduction
$f \leqW \mathsf{w}(\mathsf{c}(f))$ can project away the first component of the
output of $\mathsf{w}(\mathsf{c}(f))$ while the backward part
of the converse can be the identity.
\end{qroof}

We can note that in the proof above did not use computability aside
from the fact type-2 computable maps correspond to morphisms in the
category in question. As partial continuous maps are
all representable in $\KVPCA$, \Cref{thm:mainthmweic0} can be proven in the same
way.

\subsection{Extended Weihrauch degrees and containers over partitioned assemblies}

The notion of extended Weihrauch predicates defined by Bauer~\cite{Bauer22} is
a specialization of his notion of instance reducibility to realizability toposes
built from filter-PCAs. The rationale is that, in the Kleene-Vesley
topos~\cite[\S 4.5]{vanOosten}, a Weihrauch problem $P$ can be translated to
predicates $P'(i)$ (over some object of the topos, which corresponds to the
inputs of $P$) such that
$\forall i. \exists j. Q'(j) \Rightarrow P'(i)$ holds
if and only if $P$ is Weihrauch-reducible to $Q$. This is essentially because,
since we are in a realizability topos, the truth of this statement is
necessarily witnessed by a program which encodes the forward part of a
reduction due to the $\forall\exists$ and a backwards part due to the $\Rightarrow$
part.

By unwinding the definition, the notion can be stated elementarily from the
definition of a filter-PCA, so let us do that, assuming
an arbitrary filter-PCA $(\bA', \bA)$ is fixed for the rest of the subsection.

\begin{definition}[{\cite[Definition 3.7]{Bauer22}, see also~\cite[Definition 5.7]{KiharaLT24}}]
An \emph{extended Weihrauch predicate} is a map $p : \bA \to \pow(\pow(\bA))$.
Its \emph{support} is the set $\suppewp{p} = \{r \in \bA \mid p(r) \neq \emptyset\}$.
$p$ is said to be reducible to another extended Weihrauch predicate $q$ when
there exists codes $e_{\mathrm{fwd}}, e_{\mathrm{bwd}} \in \bA'$ such that,
for every $r \in \suppewp{p}$:
\begin{enumerate}
  \item $e_{\mathrm{fwd}} \cdot r$ is defined and belongs to $\suppewp{q}$
  \item \label{enumitem:BauerExtRedBwd} for every $\theta \in p(r)$ there is
$\xi \in q(e_{\mathrm{fwd}} \cdot r)$ such that $e_{\mathrm{bwd}} \cdot r \cdot s$
is defined and belongs to $\theta$ for any $s \in \xi$.
\end{enumerate}
\end{definition}

For the sequel, let us
restate~\ref{enumitem:BauerExtRedBwd}, using the axiom of choice, as follows:
\begin{enumerate}
  \item[\ref{enumitem:BauerExtRedBwd}'] \label{enumitem:ExtRedBwdFun}
    we have a function $f_r : p(r) \to q(e_{\mathrm{fwd}} \cdot r)$ such that
    $e_{\mathrm{bwd}} \cdot r \cdot s$
    is defined and belongs to $\theta$ for any $s \in f(\theta)$.
\end{enumerate}
With this version, it is helpful to think of $(e_{\mathrm{fwd}}, (f_r)_{r\in \suppewp{p}})$
as a code for a forward reduction and $e_{\mathrm{bwd}}$ as a code for
a backward reduction.
In the case of extended Weihrauch predicate $p$ coming from Weihrauch problems
(or slightly extended Weihrauch problems), $p(r)$ will be always a
(sub)singleton. In the general case, $p$ corresponds to a binary partial
multivalued function $\bA \times \bA \mto \bA$ where the first input corresponds
to $r$, the second to an element of $p(r)$ and the forward part of the reduction
is allowed to be an arbitrary set-theoretic map while the backward reduction
cannot compute at all with the second input.

\begin{example}[See~\cite{Bauer22}]
  \label{ex:wlem}
An example of an extended Weihrauch predicate which does not match any
Weihrauch degree
is $\mathsf{WLEM}$ which is formally defined as the constant function
mapping any $r \in \bA$ to $\{\{\underline{0}\}, \{\underline{1}\}\}$,
where $\underline{i}$ is the canonical code for the number $i$ in $\bA'$.

Assuming the axiom of choice, it is maximal for reducibility among all extended
Weihrauch predicates $p$ with $\bigcup p(r) \subseteq
\{\underline{0}, \underline{1}\}$; clearly there is no such maximal Weihrauch
problem.
\end{example}

With this definition, let us show the following theorem, which will justify
calling objects of $\pAsm(\bA',\bA)$ \emph{extended Weihrauch problems}.

\mainthmweiext*

Note that the proof, as well as the rest of the section, uses liberally the
axiom of choice.

\begin{qroof}
First let us explain how to turn an extended Weihrauch predicate $p$ into a
container $\widehat{p}$. We first define the assemblies $I_p$ and $X_p$ of
positions and directions of $\widehat{p}$ by taking
  \[
\begin{array}{lcl !\qquad rcl}
  \carrier{X_p} &=& \{(r, \theta, s) \in \bA \times \pow(\bA) \times \bA \mid r \in \suppewp{p}, \theta \in p(r), s \in \theta\} 
  & \tuple{r, s} &\Vdash_{X_p}& (r, \theta, s) \\
  \carrier{I_p} &=& \{(r, \theta) \in \bA \times \pow(\bA) \mid r \in \suppewp{p}, \theta \in p(r)\} 
  & r &\Vdash_{I_p}& (r, \theta) \\
\end{array}\]
It is obvious those assemblies are partitioned. $\widehat{p}$ is then defined
by the projection $(r, \theta, s) \mapsto (r, \theta)$, which is tracked by
code of the first projection in $\bA'$.
  Furthermore, a reduction $((e_{\mathrm{fwd}}, f), e_{\mathrm{bwd}})$
between extended Weihrauch predicates $p$ and $q$ induce a map $\widehat{p} \to \widehat{q}$
  in $\Container(\pAsm(\bA',\bA))$: the forward map $(r, \theta) \mapsto (e_{\mathrm{fwd}} \cdot r, f(\theta))$
is tracked by $e_{\mathrm{fwd}}$ and the backwards map
$(r,\theta), (e_{\mathrm{fwd}} \cdot r, f(\theta), s) \mapsto (r, \theta, e_{\mathrm{bwd}} \cdot r \cdot s)$
is tracked by $\lambda \tuple{r, \tuple{t, s}}. \; e_{\mathrm{bwd}} \cdot r \cdot s$.

Conversely, given a container $P : X \to I$ in $\Container(\pAsm(\bA',\bA))$,
  define the extended Weihrauch predicate
\[
  \begin{array}{lcl}
    \phi_P(r) &=& \{ \Phi_P(i)\mid r \Vdash_I i\}\\
    & & \text{where } \Phi_P(i) = \{ e \in \bA \mid \exists s \in P^{-1}(i). \; e \Vdash_X s\}  \\
  \end{array}
  \]
Using the axiom of choice, let us fix a section $\zeta_{P, r}$
of the restriction of $\Phi_P$ to the $i \in I$ with $r \Vdash_I i$.

If we have a map $(\varphi, \psi)$ from $P$ to $Q : Y \to J$ in $\Container(\pAsm(\bA', \bA))$,
with $e_\varphi \Vdash_{I \to J} \varphi$ and $e_\psi$ tracks $\psi$,
we can turn this into a reduction between extended predicates by taking the
forward part to be $(e_\varphi, (\varphi_r')_{r \in \suppewp{\phi_P}})$,
where $\varphi'_r(X)$ is defined by taking
\[
  \begin{array}{llcl}
    \varphi'_r :& \suppewp{\phi_P}(r) & \longto & \suppewp{\phi_Q}(e_\varphi \cdot r) \\
    & X &\longmapsto & \Phi_Q(\varphi(\zeta_{P,r}(X)) \\
  \end{array}
\]
Assuming without loss of generality that the domain
of $\psi$ is the assembly where codes are pairs of codes for $I$ and $Y$,
we can take the code $\lambda x. \lambda y. \; e_\psi \cdot \tuple{x, y}$ as
the code for the backwards part and check that this gives us a reduction from
$\phi_P$ to $\phi_Q$.

Now let us show that $\phi_{\widehat{p}}$ is always equivalent to $p$ by
exhibiting reductions both ways.
First, note that
$\suppewp{\phi_{\widehat{p}}} =
\{ r \mid \exists \theta \in p(r). \; r \Vdash_{I_p} (r, \theta) \} = \suppewp{p}$
and that $\Phi_{\widehat{p}}(r, \theta) = \{ \tuple{r, s} \in \bA \mid s \in \theta\}$.
For the reduction from $\phi_{\widehat{p}}$ to $p$, we can take
the code for the identity together with the map $\xi \mapsto \{ \pi_2 \cdot e \mid e \in \xi\}$
and the code $\lambda r. \lambda s. \tuple{r,s}$ for the backward part.
For the other direction, the forward part is obtained by taking the code for
the identity together with the $r$-indexed family of maps $s \mapsto \tuple{r , s}$,
and the backward part by $\lambda x.\lambda y. \; \pi_2 \; y$.

Finally, we can also show that there exists morphisms both ways between $P : X \to I$ and
$\widehat{\phi_P}$ in $\Container(\pAsm(\bA',\bA))$. First, note that
the carrier of the object $I'$ of positions of $\widehat{\phi_P}$ is the set of all pairs
$(r, \Phi_P(i))$ with $r \Vdash_{I} i$.
There is an obvious (surjective) map $I \to I'$
taking $i$ to the unique pair $(r, \Phi_P(i))$ such that $r \Vdash_I i$, which
is tracked by $\lambda x. \; x$. Thanks to the axiom of choice, this map has a
section $s : I' \to I$ which is also tracked by $\lambda x. \; x$. These two maps will be
the forward part of the reductions we want.

For the backward part, first note the carrier $X'$ of the object of directions of
$\widehat{\phi_P}$ is the set of all triples $(r, \Phi_P(i), e)$ with
$r \Vdash_I i$ and $e \in \Phi_P(i)$, that is those $e$ tracking some element of
$x \in X$ such $P(x) = i$, and such a triple is tracked by the pair
$\tuple{r, e}$.
Since $P$ is a map in $\pAsm(\bA',\bA)$, we have that $e$ determines the value
of all such triples. Further, $P$ is tracked by some $e_P \in \bA'$,
so we have that $e_P \cdot e = r$ for every $(r, \Phi_P(i), e) \in \carrier{X'}$.
So $\lambda x. \; \tuple{ e_P \cdot x, x}$ tracks (and determines) a map
$X \to X'$ which can be used to complete the reduction from $\widehat{\phi_P}$ to $P$.

In the other direction, let us appeal to the axiom of choice to get
functions $\epsilon_{P,r,i}$ such that
$e \Vdash_X \epsilon_{P,r,i}(e)$ and
$P(x) = i$ for every $(r, \Phi_P(i), e) \in \carrier{X'}$. We can use that
to define the map
\[
\begin{array}{lcl}
  X' &\longto& X\\
  (r, \theta, e) &\longmapsto& \epsilon_{P,r,s(r,\theta)}(e)
\end{array}
\]
which is a morphism $X' \to X$ in $\pAsm(\bA', \bA)$ tracked by
$\lambda x. \; x$ and can similarly be used to complete the reduction
from $P$ to $\widehat{\phi_P}$.
\end{qroof}

\begin{example}
The Weihrauch predicate $\mathsf{WLEM}$ translates to a container isomorphic to
$2 \to \nabla(2)$, where $\nabla : \Set \to \pAsm(\bA', \bA)$
is the right adjoint to the forgetful functor $\pAsm(\bA',\bA) \to \Set$ sending
a partitioned assembly to its carrier. It sends a set $X$ to the partitioned
assembly with carrier $X$ and realizability relation induced by $\underline{0} \Vdash_{\nabla(X)} x$
(so all elements share the same code).
The underlying map $2 \to \nabla(2)$ between carriers is the identity and is
tracked by $\lambda x. \; \underline{0}$.
\end{example}

\begin{remark}
One may note that the definition of extended Weihrauch
predicates and reduction between those comports a certain asymmetry, where
the input as well as the forward reduction is split between a computable and
non-computable part while the output and backward reduction do not
contain any non-computable component.
This contrasts the notion of extended Weihrauch problem, where inputs and
outputs are treated more uniformly as partitioned assemblies. In this setting,
the computational part consist of valid codes for the partitioned assemblies,
while the non-computational part consists of the elements of the carrier tracked
by the codes, so in principle we could have extended Weihrauch problems with
``non-computational outputs''.

This does not contradict~\Cref{thm:mainthmweiext} because we can essentially
turn any extended Weihrauch problem $P : X \to U$ into an equivalent (but not isomorphic!)
one where the non-computational is suppressed, assuming the axiom of choice:
one can quotient the carrier of $X$ by taking $x$ and $x'$ to be equivalent if
$P(x) = P(x')$ and they are tracked by the same code. Then we obtain a
partitioned assembly $X'$ and a problem $X' \to U$ which turns out to be
equivalent provided that we have a section to the quotient map $X \to X'$; this
section always exists if we assume the axiom of choice.
\end{remark}

\section{Operators on containers and Weihrauch problems}
\label{sec:structure}

In this section, we list a few classical algebraic operators that appear naturally
under some form both in the literature on the Weihrauch lattice~\cite{paulybrattka4}
and containers~\cite{polybook}. Those operators will typically correspond to (quasi-)functors
over containers, implying in particular that they are monotonous over degrees.
Those functors can also be generically defined without reference to the specifics
of type-2 computability; thus we will define them over $\Container(\cC)$, making
minimal assumptions about $\cC$ while staying consistent with $\pMod(\bA',\bA)$ and $\pAsm(\bA',\bA)$.

\subsection{The commutative tensor products}

One basic fact about Weihrauch degrees is that they form a distributive
lattice. This is reflects the fact that under some mild assumptions,
$\Container(\cC)$ has products and coproducts that distribute over one another.

\begin{definition}[\cite{carboni93extensive}]
A category with finite sums and products is called \emph{extensive}
when the canonical functor $+ : \slice{\cC}{A} \times \slice{\cC}{B} \to
\slice{\cC}{(A + B)}$ is an isomorphism.
We say a category is \emph{lextensive} if it has
all finite limits, all finite coproducts and is extensive.
\end{definition}

\begin{proposition}[{See also~\cite[(3), (6) and (7)]{spivak2024reference} for a particular case}]
\label{prop:prod-coprod}
If $\cC$ is lextensive,
$\Container(\cC)$ has finite products, coproducts and they distribute.
\end{proposition}
\begin{qroof}
It is easy to check that the maps $\id_0 : 0 \to 0$ and $!_0 : 0 \to 1$ in
$\cC$ are respectively initial and terminal objects in $\Container(\cC)$;
for the latter, this is because any pullback along any map $0 \to U$
has an initial object as domain.

Given two containers $P_1 : X_1 \to U_1$ and $P_2 : X_2 \to U_2$, a coproduct
$P_1 + P_2 : X_1 + X_2 \to U_1 + U_2$ is obtained by functoriality of $+ : \cC^2 \to \cC$.
The coprojections are horizontal morphisms represented $(\coproj_i, \id_{X_i})$;
note that $X_i$ is only a valid domain for the backward map because of extensivity.
The copairing $\cotuple{\Psi_1, \Psi_2} : X_1 + X_2 \to U$ of two reductions,
$\Psi_1 = (\phi_1, \psi_1) : X_1 \to U$ and $\Psi_2 = (\phi_2, \psi_2) : X_1
\to U$, is represented by $(\cotuple{\phi_1, \phi_2}, \psi_1 +
\psi_2)$. This backwards map has the correct domain since, in an
extensive category, the pullback of a coproduct is the coproduct of
the respective pullbacks along the same morphism.

A product of $P_1$ and $P_2$ can be defined as the map
$\cotuple{P_1 \times \id_{U_2}, \id_{U_1} \times P_2} :
  X_1 \times U_2 + U_1 \times X_2 \to U_1 \times U_2$.
The $i$-th projection $P_1 \times P_2 \to P_i$ in $\Container(\cC)$ is
represented by $(\pi_i, \coproj_i)$; again, this is only a valid
morphism because $\cC$ is lextensive. The pairing $\tuple{\Psi_1, \Psi_2} :
X \to U_1 \times U_2$ of two reductions $\Psi_1 = (\phi_1, \psi_1) : X \to
U_1$ and $\Psi_2 = (\phi_2, \psi_2) : X \to U_2$, is represented by
$(\tuple{\phi_1, \phi_2}, \cotuple{\psi_1, \psi_2})$. To see that the
domain of the backwards map works, we first note if we pull a map $P_1$ back
along $\phi_1$, the same object (with different projections) is a
pullback of $P_1 \times \id_{U_2}$ along $\tuple{\phi_1, \phi_2}$. We can
then invoke lextensivity to handle the coproduct of pullbacks.

To see that $\Container(\cC)$ is distributive, it is enough to notice
that the canonical morphism $(P_1 \times Q) + (P_2 \times Q) \to (P_1 + P_2) \times Q$ is
represented by a horizontal morphism $(c, \id)$ where $c$ is the
canonical isomorphism in $\cC$ witnessing distributivity for the base.
\end{qroof}

\begin{proposition}
Assuming $\cC$ is lextensive, answerable containers of $\Container(\cC)$ are
stable under binary products and coproducts. The initial object is an answerable container but
not the terminal object.
\end{proposition}
\begin{qroof}
For binary products, we first note that $P_1 \times P_2$
in $\Container(\cC)$ is obtained by copairing two pullback-stable epimorphisms:
we can show that $P_1 \times \id_{U_2}$ is
a pullback-stable epimorphism since it is the
pullback of $P_1$ along the projection $\pi_1 : U_1 \times U_2 \to U_1$, and
similarly for $\id_{U_1} \times P_2$.
Now, in extensive categories, the pullback of a copairing is
the copairing of the pullbacks, and a copairing is epimorphic as long as
one of its component is an epi, so we can conclude.

$P_1 + P_2$ being represented by a pullback-stable epi follows from the fact that
coproducts are disjoint in an extensive category. Extensivity also ensures
that if we have a map $X \to 0$ in $\cC$, $X$ is initial; hence, $\id_0 : 0 \to 0$
is a pullback-stable epimorphism.
\end{qroof}

Containers also admit another natural monoidal product, which is sometimes
called the \emph{parallel product}. The parallel product of $P_1 : X_1 \to U_1$
and $P_2 : X_2 \to U_2$, which we write $P_1 \tensor P_2$, is represented by
the morphism $P_1 \times P_2 : X_1 \times X_2 \to U_1 \times U_2$ obtained by
the functoriality of $\times$ in $\cC$
(assuming $\cC$ has cartesian products)\cite[Definition 3.65]{polybook}.
Regarded as a Weihrauch problem, this means we may ask a question to both
$P_1$ and $P_2$ and get back answers for both questions.

It is straightforward to check that
$\tensor$ is a symmetric monoidal product in $\cC$
(see~\cite[Chapter VII]{mac2013categories}) and that answerable containers are
stable under $\tensor$. If we further assume that $\cC$
is lextensive, we can show that there are further distributive laws.

\begin{proposition}[{See also~\cite[(6) and (28)]{spivak2024reference} for a particular case}]
If $\cC$ is lextensive, then we have natural transformations
  $P \tensor Q + P \tensor R ~\cong~ P \tensor (Q + R)$ and
$(P \tensor Q) \times R \to P \tensor (Q \times R)$ in $\Container(\cC)$.
\end{proposition}
\begin{qroof}
The proof that $\tensor$ distributes over $+$ is essentially the same
as the proof that $\times$ distributes, so we omit it.

For the second claim, assuming that we have $P : X \to U$, $Q : Y \to V$ and $R : Z \to W$,
the natural transformation is represented by the vertical map
$(\id, c)$ with $c = \id_{X \times Y \times W} + (P \times \id_{V \times Z}) : 
(X \times Y \times W) + (X \times V \times Z) \to (X \times Y \times W) +
(U \times V \times Z)$. Note that it is a valid morphism representative only because
the domain of $c$ is isomorphic to $X \times (Y \times W + V \times Z)$;
this is true because $\cC$ is distributive.
\end{qroof}

Note that with this and the functoriality of $\times$, when moving to
$\posetOf{\Container(\cC)}$ we recover all of the
axioms listed in~\cite{theoryWeiTimes} pertaining to binary joins, meets and
parallel products.

\subsection{The composition product in nicer categories of containers}

Another monoidal product which appears in the literature on containers is the
\emph{composition product}. The most natural way to introduce this is maybe by
making the link between containers and polynomial
functors~\cite[Theorem 2.12]{GKpoly}.
In short, assuming that $\cC$ is locally cartesian closed, we map
any container $P : X \to U$ to the endofunctor over $\cC$ defined by
$\Sigma_{!_U} \circ \Pi_P \circ {!^*_X} : \slice{\cC}{1} \to \slice{\cC}{1}$
and the isomorphism $\dom : \slice{\cC}{1} \cong \cC$.
Regarding $P$ as an internal family $(X_u)_{u \in U}$ to $\cC$,
on objects this functor maps $A$ to $\sum_{u \in U} A^{X_u}$.
Those endofunctors are called
\emph{polynomial endofunctors} over $\cC$, and one important fact is that
strong natural transformations between those correspond exactly to
container morphisms.
One important theorem is that polynomial endofunctors are stable under
composition, which induces a monoidal product on polynomials sometimes called the
\emph{substitution} or
\emph{sequential product}~\cite[(9) Proposition 1.12]{GKpoly}.

In the spirit of Weihrauch reducibility, the sequential product $P \star Q$ can be regarded
as the problem whose inputs are composed of an input $v$ for $Q$ together with
a function $f$ turning a solution for said input into an input for $P$. A solution
for $P \star Q$ then consists of a solution for $v$ and a solution for $f(v)$.
In sets, assuming we have $Q : Y \to V$, $X_u = P^{-1}(u)$ and $Y_v = Q^{-1}(v)$ for every
$u, v$, $P \star Q$
is the family $\left(\sum_{y \in Y_v} X_{f(y)}\right)_{(v,f)}$
whose indexing set is $\sum_{v \in V} \left(Y_v \to X\right)$. This can be internalized
in any locally cartesian closed category (see \Cref{fig:compositionProduct}).

In short, reducing to $P \star Q$ is the ability to make an oracle call to $Q$
and then $P$.
The sequential product of
Weihrauch degrees is designed to capture this intuition and is thus
defined this way~\cite[Definition 3]{westrick2020}\footnote{The
original definition in~\cite{paulybrattka4} actually has a slightly different flavour,
but they correspond to the same Weihrauch degree.}, but for a small wrinkle: $\pMod(\KVPCAR,\KVPCA)$
is not locally cartesian closed!

\begin{figure}
\[\begin{tikzcd}[column sep=2.25em]
	{\sum\limits_{v \in V} Y_v \times \sum\limits_{f \in U^{Y_v}} X_{f(v)}} & {\sum\limits_{v \in V} Y_v \times U^{Y_v}} & {\sum\limits_{v \in V} Y_v \times U^{Y_v}} & {\sum\limits_{v \in V}U^{Y_v}} \\
	{X \times Y} & {U \times Y} \\
	X & U & Y & V
	\arrow[from=1-1, to=1-2]
	\arrow["{P \star Q}", curve={height=-24pt}, from=1-1, to=1-4]
	\arrow[from=1-1, to=2-1]
	\arrow["\lrcorner"{anchor=center, pos=0.125}, draw=none, from=1-1, to=2-2]
	\arrow[shift left, no head, from=1-2, to=1-3]
	\arrow[shift right, no head, from=1-2, to=1-3]
    \arrow["\varepsilon_{\pi_2}", from=1-2, to=2-2]
	\arrow[from=1-3, to=1-4]
	\arrow[from=1-3, to=3-3]
	\arrow["\lrcorner"{anchor=center, pos=0.125}, draw=none, from=1-3, to=3-4]
	\arrow["{\Pi_Q(\pi_2)}", from=1-4, to=3-4]
	\arrow["{P \times \id}", from=2-1, to=2-2]
	\arrow["{\pi_1}"', from=2-1, to=3-1]
	\arrow["\lrcorner"{anchor=center, pos=0.125}, draw=none, from=2-1, to=3-2]
	\arrow["{\pi_1}", from=2-2, to=3-2]
	\arrow["{\pi_2}", from=2-2, to=3-3]
	\arrow["P"', from=3-1, to=3-2]
	\arrow["Q"', from=3-3, to=3-4]
\end{tikzcd}\]
\caption{Diagrammatic definition of the composition product $P \star Q$ of
  two containers $P$ and $Q$ (simplified version of~\cite[(9)]{GKpoly}). To ease reading, we name objects in the top row
according to their definition in type theory; the definition is determined
(up to isomorphism) by $P$, $Q$, the projections out of $U \times Y$, $\Pi_Q(\pi_2)$, the denoted pullback
squares and $\varepsilon_{\pi_2}$ being to the counit of the adjunction $Q^* \dashv \Pi_Q$ at $\pi_2$.
  $P \star Q$, when regarded as a polynomial functor $\interp{P \star Q}$, corresponds to the
  reverse composition $\interp{Q} \circ \interp{P}$ of functors.}
  \label{fig:compositionProduct}
\end{figure}

\begin{proposition}
  \label{prop:pmod-notccc}
$\pMod(\KVPCA)$, $\pAsm(\KVPCA)$, $\pMod(\KVPCAR, \KVPCA)$
and $\pAsm(\KVPCAR, \KVPCA)$ are not cartesian closed.
\end{proposition}
\begin{qroof}
Let us focus on $\pAsm(\KVPCA)$ (proofs along the same lines work for the other
categories). We are going to show there is no exponential
object for $2^{\Baire}$; let us call $E$ such an exponential object, which
comes with an evaluation map $E \times \Baire \to 2$ and a universal property.
First, because of the uniqueness part of the universal property, we must have
that $E$ is modest.
We can consider the partitioned modest set $T$ of well-founded trees with
countable branching whose leaves are labeled in $2$. There is a computable map
$T \times \Baire \to 2$ that assigns each pair $(t, p)$ to the unique boolean
that the path $p$ traverses in $t$, so, using the exponential structure, we
have a fortiori a surjective map $r : T \to E$ in $\pMod(\KVPCA)$. Using the
evaluation map, we can continuously map a code of $E$ into a code tracking the
underlying function $f : \Baire \to 2$, and from such a code is up to encoding
details an element of $T$ encoding $f$. Hence we have a section $s$ of $r$.

Now consider some $t$ which is in the image of $s$, and replace one of its
leaf $l$ by an infinite well-founded tree whose leaves are labelled the same way
to obtain $t'$. We have $s(r(t')) = t$. But since $s \circ r$ is continuous,
it must only inspect a finite part of $t'$ to determine the label of $l$.
Then considering $t''$ obtained from $t'$ by flipping one of the unexamined leaves
that did not belong to $t$ leads to a contradiction.
\end{qroof}

The failure of cartesian closure is essentially due
to a lack of quotients rather than a lack of power: it is not that we cannot
build modest sets of codes for function spaces, it is that we cannot guarantee
that a function has a unique code.
It means in particular that we still have a weak version of local cartesian closure for
$\pMod$, which is enough to talk about spaces of codes for functions; this is
why the same intuition can be used to build the sequential product of Weihrauch
problems.
But a disadvantage is that we then need to reconstruct from scratch certain
proofs, for instance that the sequential product on the left distributes over
coproducts. In the next two subsections, we rather present a strategy to
derive those results by observing that $\pMod$ is the full subcategory of
regular projective objects of the locally-cartesian closed category $\Mod$ and
that $\Mod$ has enough projectives.

\subsection{Weak structure, projectives and quasi-functors}

The definition of various canonically determined structures like
cartesian closure comes from the characterizations of \emph{adjoint functors}.
One of the shortest definition states that a functor $L : \cC \to \cD$ is
a left adjoint to the functor $R : \cD \to \cC$ if we have isomorphisms
\[ \Hom{\cD}{L(X)}{A} \cong \Hom{\cC}{X}{R(A)} \qquad \text{natural in $A$ and $B$}\]
To characterize weak structure, one can take a similar approach. However, this
requires weakening the notion of what is a right adjoint.

\begin{definition}[{\cite[\S 1]{kainen1971weak}}]
A weak right adjoint of the functor $F : \cC \to \cD$ consists of
\begin{itemize}
\item a map $G$ from objects of $\cD$ to objects of $\cC$
\item and natural surjections $\Hom{\cC}{-}{G(A)} \epito \Hom{\cD}{F(-)}{A}$.
\end{itemize}
\end{definition}

\begin{example}
For any small category $\cC$, the category $\cC \to \posetOf{\cC}$ has
an obvious weak right adjoint if we assume the axiom of choice.
\end{example}

Analogously to the definition of locally-cartesian closed categories, we
say that a category is \emph{weakly} locally cartesian closed when all functors
$f^*$ have weak right adjoints~\cite[Remark 3.2]{CR00}.

\begin{definition}
  \label{def:regularepi}
  A \emph{regular epimorphism} is an epimorphism which is a coequalizer of some
pair of morphisms.
\end{definition}

\begin{example}
In topological spaces, and a fortiori most of the subcategories of those that we
consider in this paper, the monomorphisms $2 \monoto \Sierp$ are epimorphic, pullback-stable
but not regular.
\end{example}

A characterization of regular epimorphisms in (multi)-represented spaces is that they
have a multi-valued inverse. In the language of assemblies, this translates to
the following lemma.

\begin{proposition}[{\cite[Proof of Theorem 1.5.2]{vanOosten}}] 
\label{prop:regularepi}
Both in $\Mod(\bA',\bA)$ and $\Asm(\bA', \bA)$, a morphism $f : X \to Y$
tracked by $e \in \bA'$ is
 a regular epimorphism if and only if there is $e' \in \bA'$ such that
\begin{itemize}
  \item there exists $x$ such that $e' \cdot e'' \Vdash_X x$ for every $y$ and $e'' \Vdash_Y y$
\item $\lambda x. \; e \cdot (e' \cdot x)$ tracks the identity on $Y$
\end{itemize}
\end{proposition}

\begin{definition}
\label{def:regularprojectives}

An object $X$ is called \emph{(regular) projective} when it has the left lifting
property against regular epimorphisms, that is, whenever we have a regular
epimorphism $e : Z \epito Y$ and a map $f : X \to Y$, there exists
some $g : X \to Z$ such that $f = e \circ g$.
\[\begin{tikzcd}
	& Z \\
	X & Y
	\arrow["e", two heads, from=1-2, to=2-2]
	\arrow[dotted, from=2-1, to=1-2]
	\arrow["f"', from=2-1, to=2-2]
\end{tikzcd}\]

For a category $\cC$, we call $p\cC$ its full subcategory consisting of regular
projective objects.

A \emph{(regular) projective cover} of an object $X$ is a projective object
$\widetilde{X}$ together with a regular epimorphism $\widetilde{X} \epito X$
\end{definition}

Turning back to the specific categories of interest here,
for an assembly $X$ of $\pAsm(\bA',\bA)$, a projective cover
$\widetilde{X}$ can be defined as follows: the carrier is
$\carrier{\widetilde{X}} = \Vdash_X \subseteq \bA \times X$,
i.e. the set of all pairs $(e, x)$ such that $e$ realizes $x$, and realizability
relation is $e \Vdash_{\widetilde{X}} (e,x)$. The same construction works for
modest sets, where we can simplify things further by simply restricting the
carrier to its first component, since modesty means that a code uniquely
determines the elements we want to code. So all in all, we have the following.

\begin{lemma}
  \label{lem:pmod-pass-projectives}
The projective objects of modest sets (assemblies) are, up to isomorphism, the
partitioned modest sets (assemblies). Furthermore, the categories
of modest sets and assemblies always have enough projectives.
\end{lemma}
\begin{qroof}
For modest sets, this is~\cite[Theorem 3.1]{BauerPhD}. For assemblies, we can
deduce that from the characterization of partitioned assemblies as the
projectives of the realizability topos
$\mathsf{RT}(\bA',\bA)$~\cite[Proposition 3.2.7]{vanOosten} by noting that
  coequalizers are preserved by the embedding $\Asm(\bA', \bA) \to \mathsf{RT}(\bA', \bA)$.
\end{qroof}

Categories $\cC$ with enough projectives
\emph{almost} have a coreflection on their projectives, which we can
characterize via a weak right adjoint.

\begin{lemma}
  \label{lem:proj-weakrightadjoint}
If $\cC$ has enough regular projectives, the full and faithful embedding induced by
the inclusion $p\cC \subseteq \cC$
has a weak right adjoint $X \mapsto \widetilde{X}$ with
$\widetilde{X} = X$ for any object $X$ of $p\cC$.
\end{lemma}
\begin{qroof}
For any object $X$, fix a regular projective cover
$\varepsilon_X : \widetilde{X} \epito X$, taking the identity when $X$ was already
regular projective. This gives a counit for our weak adjoint, whose
morphism part is then the postcomposition
\[\varepsilon_X \circ - : \Hom{p\cC}{A}{\widetilde{X}} \epito \Hom{\cC}{A}{X}\]
It is obviously natural in $A$. It is easily seen to be surjective since $A$ is
regular projective and $\varepsilon_X$ is a regular epimorphism.
\end{qroof}

\begin{remark}
At this stage, one can use that functor to prove that $p\cC$ is weakly
locally cartesian closed from the fact that $\cC$ is locally cartesian closed
and has enough projectives. In fact this can be strengthened to an equivalence~\cite[Theorem 3.3]{CR00}.
\end{remark}

Something to be noted is that weak right adjoints need not be functors; however
they are still \emph{quasi-functors} in the following sense.

\begin{definition}[{\cite[\S 0]{kainen1971weak}}]
A quasi-functor $F : \cC \leadsto \cD$ maps objects of $\cC$ to objects of $\cD$
  and morphisms $X \to Y$ of $\cC$ to non-empty sets of morphisms
  $F(X) \to F(Y)$ such that
  \begin{itemize}
    \item $\id_{F(X)} \in F(\id_X)$ for every object $X$ of $\cC$
    \item if $f' \in F(f)$ and $g' \in F(g)$ for composable arrows
    $f$ and $g$, then $f' \circ g' \in F(f \circ g)$.
  \end{itemize}
\end{definition}

While quasi-functors are weaker than functors, we can note that the two notions
coincide when the codomain is a preorder. Quasi-functors over $\cC$, much
like functors, also induce monotonous operations over the poset $\posetOf{\cC}$.

\subsection{The composition product for containers on projectives}

Now we will define a sequential product $\star_p$ on $\Container(p\cC)$ in
terms of the sequential product $\star$ on $\Container(\cC)$.
First, we will use the following definition to introduce a nice full subcategory
of $\Container(\cC)$.

\begin{definition}
Say that a morphism $f : X \to U$ is \emph{\fiberwiseprojective{}} if, for every
$g : V \to U$ with $V$ regular projective, then the domain of the pullback
$g^*(f)$ is also regular projective.
\end{definition}

\Fiberwiseprojective{} morphisms are easily seen to be closed under composition and
arbitrary pullbacks. Let us call $\fpContainer(\cC)$ the full subcategory of
$\Container(\cC)$ whose objects are those containers that are
\fiberwiseprojective{} when regarded as morphisms of $\cC$. 

\begin{lemma}
  \label{lem:fpContainerWeakRightAdjoint}
Suppose that $p\cC$ has finite limits which are preserved
by the inclusion $p\cC \subseteq \cC$. If $\cC$ has enough regular projectives, then
the functor induced by the inclusion $\Container(p\cC) \subseteq \fpContainer(\cC)$ has a
weak right adjoint $P \mapsto \widetilde{P}$ with $P = \widetilde{P}$ for any object $P$ of
$\Container(p\cC)$.
\end{lemma}
\begin{qroof}
For any $P : X \to U$, we can obtain $\widetilde{P}$ by pulling back along
a regular projective cover $\widetilde{U} \epito U$; since we assume $P$
was \fiberwiseprojective{}, this ensures $\widetilde{P}$ belongs to $\fpContainer$.
The pullback itself gives us an horizontal morphism
$\varepsilon_P : \widetilde{P} \to P$ in $\Container$; much like in the proof
of~\Cref{lem:proj-weakrightadjoint}, this gives us a counit for a weak right
adjoint and its morphism part is obtained by postcomposing $\varepsilon_P$.
This is surjective because, given a morphism $(\varphi, \psi)$ from $Q$
to $P$, we can obtain a morphism $(\tilde{\varphi}, \psi \circ \theta)$
from $Q$ to $\widetilde{P}$ as shown in the commuting diagram below.
A morphism $\widetilde{\varphi}$ such that $\varepsilon_U \circ \widetilde{\varphi} = \varphi$
exists because $V$ is projective and $\theta$ is defined using the universal
property of the pullback of $P$ along $\varphi$.

\[\begin{tikzcd}
	&&& \cdot && \cdot \\
	Q && \cdot && X \\
	&&& V && {\widetilde{U}} \\
	V && V && U
	\arrow[from=1-4, to=1-6]
	\arrow["\theta"', dotted, from=1-4, to=2-3]
	\arrow[from=1-4, to=3-4]
	\arrow["\lrcorner"{anchor=center, pos=0.125}, draw=none, from=1-4, to=3-6]
	\arrow[from=1-6, to=2-5]
	\arrow["{\widetilde{P}}", from=1-6, to=3-6]
	\arrow["Q", from=2-1, to=4-1]
	\arrow["\psi"', from=2-3, to=2-1]
	\arrow[from=2-3, to=2-5]
	\arrow[from=2-3, to=4-3]
	\arrow["\lrcorner"{anchor=center, pos=0.125}, draw=none, from=2-3, to=4-5]
	\arrow[""{name=0, anchor=center, inner sep=0}, "P"{pos=0.7}, from=2-5, to=4-5]
	\arrow["{{\widetilde{\varphi}}}"{pos=0.3}, dashed, from=3-4, to=3-6]
	\arrow["{{\varepsilon_U}}", two heads, from=3-6, to=4-5]
	\arrow[shift right, no head, from=4-1, to=4-3]
	\arrow[shift right, no head, from=4-3, to=3-4]
	\arrow[shift left, no head, from=4-3, to=3-4]
	\arrow[shift right, no head, from=4-3, to=4-1]
	\arrow["\varphi", from=4-3, to=4-5]
	\arrow["\lrcorner"{anchor=center, pos=0.125, rotate=-90}, draw=none, from=1-6, to=0]
\end{tikzcd}
\]
$\phantom{aa}$
\end{qroof}

Now, assuming $\cC$ is locally cartesian closed and $p\cC$ is
stable under pullback, the sequential
product $\star$ actually restricts to a functor
$\Container(p\cC) \times \Container(p\cC) \to \fpContainer(\cC)$; this is
because the map representing $P \star Q$ is the composition of two maps
obtained by pullback from $P$ and $Q$ (see~\Cref{fig:compositionProduct}).
Hence we can define a quasi-functor
$- \star_p - : \Container(p\cC) \times \Container(p\cC) \leadsto \Container(p\cC)$ by
postcomposing with the weak right adjoint $\fpContainer(\cC) \to \Container(p\cC)$
we just defined.

\begin{remark}
While the precise definition of $- \star_p -$ up to isomorphism\footnote{Up to
equality, it depends also on a choice of pullbacks.} does depend
on the choice of projective covers, the operator on the posetal reflection
$\posetOf{\Container(p\cC)}$ does not depend on that.

Furthermore, when $\cC$
has enough projectives, the weak right adjoint $\widetilde\Pi_f$ to $f^*$
actually yields projective covers of $\Pi_f$; this means that $P \star_p Q$
may coincide exactly with the construction obtained by mimicking the definition
of $P \star Q$ by replacing all instances of $\Pi_Q$ with its weak version.
\end{remark}

\begin{remark}
The fact that $\star_p$ is defined via a weak right adjoint ensures that
for any objects $P$ and $Q$ of $\Container(p\cC)$,
there is always a map $P \star_p Q \to P \star Q$ in $\fpContainer(\cC)$.
However we conjecture that for $\cC = \Mod(\KVPCAR,\KVPCA)$, there are
$P$ and $Q$ such that there are no maps $P \star Q \to P \star_p Q$.
\end{remark}

We can now use this to transfer some easy results about $\star$ to results
about $\star_p$ (and a fortiori the sequential product in the (extended)
Weihrauch degrees) in a generic way. By way of example, let us rederive some
known inequalities in the Weihrauch degrees.

\begin{lemma}[{See e.g.~\cite[(26)]{spivak2024reference}}]
If $\cC$ is locally cartesian closed and extensive, then in $\Container(\cC)$ we have
maps
$(P \star Q) \times R
\to
(P \times R) \star Q$.
\end{lemma}
\begin{qroof}
Calling $\interp{P}$ the polynomial functor corresponding to $P$, going
though the equivalence of categories with polynomial functors alluded to before,
it suffices to provide a strong natural transformation
\[\interp{(P \star Q) \times R} \quad\Longrightarrow\quad \interp{(P \times R) \star Q}\]
Since we have an equivalence of categories, that the cartesian product of
(polynomial) functors are computed pointwise
in functor categories and that sequential product corresponds to (reverse) composition
of polynomial functors, it thus suffices to provide a strong natural transformation
\[(\interp{Q} \circ \interp{P}) \times \interp{R} \quad\Longrightarrow\quad \interp{Q} \circ (\interp{P} \times \interp{R})\]
For this, it is sufficient to provide a transformation which is natural in $A$ and $B$
\[ \interp{Q}(A) \times B \quad\longrightarrow \quad \interp{Q}(A \times B) \]
But this amounts to showing that $\interp{Q}$ is a strong functor, which is
the case as it is a polynomial functor~\cite[\S 1.2]{GKpoly}.
\end{qroof}

\begin{corollary}
$
(a \star b) \sqcap c \le
(a \sqcap c) \star b
$
is valid in the (extended) Weihrauch degrees.
\end{corollary}
\begin{qroof}
First note that if we have that $\cC$ is lextensive, has enough
projectives and that its projectives are closed under coproduct and pullbacks,
then $\times$ is preserved by the embedding $\Container(p\cC) \subseteq 
\fpContainer(\cC)$ preserves the cartesian product and coproducts built
in $\Container(p\cC)$. As a result, assuming now that $\cC$ is also locally
cartesian closed, we have that
$(P \star_p Q) \times R$ is obtained by applying the weak right adjoint to
$\Container(p\cC) \subseteq \fpContainer(\cC)$ to $(P \star Q) \times R$;
by definition, it is also the case for $(P \times R) \star_p Q$. Hence we can
map the morphism $(P \star Q) \times R \to (P \times R) \star Q$
through the weak right adjoint and obtain a morphism
that witnesses the validity of the inequality in $\posetOf{\Container(p\cC)}$.
The corollary can then be derived since
$\Asm(\KVPCAR,\KVPCA)$ and $\Mod(\KVPCAR, \KVPCA)$ are extensive and locally cartesian closed.
\end{qroof}

While this approach can probably also help derive further distributive laws and
similar results on other operators
that also rely on the weak local cartesian closure to be defined, such as 
various closures on slightly extended Weihrauch degrees\footnote{
In the literature on Weihrauch degrees, the monoidal closure,
that is the right adjoint to $Q \mapsto Q \tensor P$, has
recently been studied in e.g.~\cite{almmv,kmp20,MVRamsey}. The cartesian
closure has been shown not to exist on the Weihrauch
degrees~\cite[Theorem 4.9]{paulykojiro}; but it does exist in the slightly
extended degrees~\cite{containerscc}.
The left adjoint to $Q \mapsto P \star Q$ has been studied a bit more
systematically and was introduced formally in~\cite[\S 3.3]{paulybrattka4}.
All of those operators appear in the extended literature on containers -
see e.g.~\cite{spivak2024reference} for their definition in $\Container(\Set)$.
},
it is not yet clear to us if it is powerful enough to allow to derive all
trivial inequalities without introspecting the definitions further.
For instance, it is not clear to us how to derive that $\star$ is
associative in the Weihrauch degrees just using that $\star$ is
a monoidal product in $\fpContainer(\Asm(\KVPCAR,\KVPCA))$
and that $\star_p$ is
$\star$ postcomposed with a quasi-functor $\fpContainer(\Asm(\KVPCAR,\KVPCA))
\leadsto \Container(\pAsm(\KVPCAR,\KVPCA))$.

\section{Conclusion}

We hope that making the connection explicit between Weihrauch reducibility and
containers will allow to connect more tightly the literature on the two topics.
For instance, it would be nice to relate formally
universal equational theories of the Weihrauch lattice with operators, like the
ones studied in~\cite{theoryWeiTimes,pradic25}, to similar theories interpreted in categories
of containers over toposes with $\cW/\cM$-types. Also, since we have shown that
Weihrauch problems are a definable subclasses of containers in assemblies over
$(\KVPCAR, \KVPCA)$, it might be a way to formalize the theory of Weihrauch reducibility in 
in proof assistants in a synthetic style (see e.g.~\cite{ForsterKM23} for similar work
for Turing-reducibility in Rocq and~\cite{KPT22} for a development in a synthetic style
in $\Asm(\KVPCAR,\KVPCA)$).

Another area which might be worth investigating is the connections between
Weihrauch reducibility and proof theory, as one can naturally interpret linear
type theories in fibrations that are closely linked to categories of
containers~\cite{glehnmoss18}. Could this be exploited to characterize
Weihrauch reducibility in a type theory, a bit in the spirit of~\cite{UFTRING20,Yoshimura2}?
One difficulty with a naive attempt is that interpreting types directly in
$\mathfrak{Fam}(\mathfrak{Fam}^\op(\id_{\pAsm(\KVPCAR,\KVPCA)}))$
(see~\cite{hofstra11mlhc} for a definition) seems somewhat counter-intuitive,
since $\mathbf{I} \leqW P \multimap Q$ holds iff $P$ reduces to (``is less powerful than'') $Q$.

\begin{credits}
\subsubsection{\ackname} Many thanks to Eike Neumann, Arno Pauly and Manlio
Valenti for discussions (especially on \Cref{prop:pmod-notccc}) and encouragements.
Many thanks as well to Takayuki Kihara and an anonymous reviewer for pointers to
the literature and comments that helped improve the presentation of the paper.

The authors acknowledge support by Swansea University and grant to Swansea
University a non-exclusive, irrevocable, sub-licensable, worldwide license to make
the accepted manuscript available on its institutional repository.
\end{credits}

\bibliographystyle{splncs04}
\bibliography{bi}

\begin{thebibliography}{10}
\providecommand{\url}[1]{\texttt{#1}}
\providecommand{\urlprefix}{URL }
\providecommand{\doi}[1]{https://doi.org/#1}

\bibitem{containers03}
Abbott, M.G., Altenkirch, T., Ghani, N.: Categories of containers. In: Gordon,
  A.D. (ed.) {FOSSACS} 2003 proceedings. Lecture Notes in Computer Science,
  vol.~2620, pp. 23--38. Springer (2003). \doi{10.1007/3-540-36576-1\_2}

\bibitem{AhmanBauer24}
Ahman, D., Bauer, A.: Comodule representations of second-order functionals
  (2024), \url{https://arxiv.org/abs/2409.17664}

\bibitem{containerscc}
Altenkirch, T., Levy, P., Staton, S.: Higher-order containers. In: Ferreira,
  F., L{\"o}we, B., Mayordomo, E., Mendes~Gomes, L. (eds.) Programs, Proofs,
  Processes. pp. 11--20. Springer Berlin Heidelberg, Berlin, Heidelberg (2010)

\bibitem{almmv}
Andrews, U., Lempp, S., Marcone, A., Miller, J.S., Valenti, M.: A jump operator
  on the {W}eihrauch degrees. arXiv 2402.13163 (2024)

\bibitem{BauerPhD}
Bauer, A.: The realizability approach to computable analysis and topology.
  Ph.D. thesis, Carnegie Mellon University (2000)

\bibitem{Bauer22}
Bauer, A.: Instance reducibility and {W}eihrauch degrees. Log. Methods Comput.
  Sci.  \textbf{18}(3) (2022). \doi{10.46298/LMCS-18(3:20)2022},
  \url{https://doi.org/10.46298/lmcs-18(3:20)2022}

\bibitem{BGMBW12}
Brattka, V., Gherardi, G., Marcone, A.: The {B}olzano–{W}eierstrass {T}heorem
  is the jump of {W}eak {K}őnig’s {L}emma. Annals of Pure and Applied Logic
  \textbf{163}(6),  623–655 (Jun 2012). \doi{10.1016/j.apal.2011.10.006}

\bibitem{BGPsurvey}
Brattka, V., Gherardi, G., Pauly, A.: {W}eihrauch complexity in computable
  analysis. CoRR  \textbf{abs/1707.03202} (2017)

\bibitem{paulybrattka4}
Brattka, V., Pauly, A.: On the algebraic structure of {W}eihrauch degrees.
  Logical Methods in Computer Science  \textbf{14}(4) (2018).
  \doi{10.23638/LMCS-14(4:4)2018}

\bibitem{CR00}
Carboni, A., Rosolini, G.: Locally cartesian closed exact completions. Journal
  of Pure and Applied Algebra  \textbf{154}(1),  103--116 (2000).
  \doi{10.1016/S0022-4049(99)00192-9}

\bibitem{carboni93extensive}
Carboni, A., Lack, S., Walters, R.: Introduction to extensive and distributive
  categories. Journal of Pure and Applied Algebra  \textbf{84}(2),  145--158
  (1993). \doi{10.1016/0022-4049(93)90035-R},
  \url{https://www.sciencedirect.com/science/article/pii/002240499390035R}

\bibitem{ForsterKM23}
Forster, Y., Kirst, D., M{\"{u}}ck, N.: Oracle computability and turing
  reducibility in the calculus of inductive constructions. In: Hur, C. (ed.)
  Programming Languages and Systems - 21st Asian Symposium, {APLAS} 2023,
  Taipei, Taiwan, November 26-29, 2023, Proceedings. Lecture Notes in Computer
  Science, vol. 14405, pp. 155--181. Springer (2023)

\bibitem{GKpoly}
Gambino, N., Kock, J.: Polynomial functors and polynomial monads. Mathematical
  Proceedings of the Cambridge Philosophical Society  \textbf{154}(1),
  153–192 (Sep 2012). \doi{10.1017/s0305004112000394}

\bibitem{paulykojiro}
Higuchi, K., Pauly, A.: The degree-structure of {W}eihrauch-reducibility.
  Logical Methods in Computer Science  \textbf{9}(2) (2013).
  \doi{10.2168/LMCS-9(2:2)2013}

\bibitem{HirschThesis90}
Hirsch, M.D.: Applications of topology to lower bound estimates in computer
  science. Ph.D. thesis, University of California, Berkeley (1990)

\bibitem{hofstra11mlhc}
Hofstra, P.J.W.: The {D}ialectica monad and its cousins. In: Makkai, M., Hart,
  B. (eds.) {Models, Logics, and Higher-dimensional Categories: A Tribute to
  the Work of Mih{\'a}ly Makkai}. CRM proceedings \& lecture notes, American
  Mathematical Society (2011)

\bibitem{iljazovic2021computability}
Iljazovi{\'c}, Z., Kihara, T.: Computability of subsets of metric spaces. In:
  Handbook of Computability and Complexity in Analysis, pp. 29--69. Springer
  (2021)

\bibitem{kainen1971weak}
Kainen, P.C.: Weak adjoint functors. Mathematische Zeitschrift  \textbf{122},
  ~1--9 (1971)

\bibitem{Kihara16BP}
Kihara, T.: Borel-piecewise continuous reducibility for uniformization
  problems. Logical Methods in Computer Science  \textbf{12}(4) (Apr 2017).
  \doi{10.2168/lmcs-12(4:4)2016}

\bibitem{KiharaLT24}
Kihara, T.: Rethinking the notion of oracle: A prequel to {L}awvere-{T}ierney
  topologies for computability theorists (2024),
  \url{https://arxiv.org/abs/2202.00188}

\bibitem{kmp20}
Kihara, T., Marcone, A., Pauly, A.: Searching for an analogue of {$\rm ATR_0$}
  in the {W}eihrauch lattice. The Journal of Symbolic Logic  \textbf{85}(3),
  1006--1043 (2020). \doi{10.1017/jsl.2020.12}

\bibitem{KPT22}
Kone{\v{c}}n{\'y}, M., Park, S., Thies, H.: Certified computation
  of nondeterministic limits. In: NASA Formal Methods. pp. 771--789. Springer
  (2022). \doi{10.1007/978-3-031-06773-0_41}

\bibitem{lmpsv}
Lempp, S., Miller, J., Pauly, A., Soskova, M., Valenti, M.: Minimal covers in
  the {W}eihrauch degrees. Proceedings of the American Mathematical Society
  \textbf{152}(11),  4893--4901 (2024)

\bibitem{mac2013categories}
Mac~Lane, S.: Categories for the working mathematician, vol.~5. Springer
  Science \& Business Media (2013)

\bibitem{MVRamsey}
Marcone, A., Valenti, M.: The open and clopen {R}amsey theorems in the
  {W}eihrauch lattice. The Journal of Symbolic Logic  \textbf{86}(1),  316--351
  (March 2021). \doi{10.1017/jsl.2021.10}

\bibitem{glehnmoss18}
Moss, S.K., von Glehn, T.: {D}ialectica models of type theory. In: Proceedings
  of the 33rd Annual {ACM/IEEE} Symposium on Logic in Computer Science, {LICS}
  2018, Oxford, UK, July 09-12, 2018. pp. 739--748 (2018)

\bibitem{theoryWeiTimes}
Neumann, E., Pauly, A., Pradic, C.: The equational theory of the {W}eihrauch
  lattice with multiplication. CoRR  \textbf{abs/2403.13975} (2024).
  \doi{10.48550/ARXIV.2403.13975},
  \url{https://doi.org/10.48550/arXiv.2403.13975}

\bibitem{polybook}
Niu, N., Spivak, D.I.: Polynomial functors: A mathematical theory of
  interaction (2024), \url{https://arxiv.org/abs/2312.00990}

\bibitem{vanoosten2016classicalrelativerealizability}
van Oosten, J., Zou, T.: Classical and relative realizability (2016),
  \url{https://arxiv.org/abs/1603.03621}

\bibitem{PaulyManyOneRedAbstract}
Pauly, A.: Many-one reductions between search problems. CoRR
  \textbf{abs/1102.3151} (2011), \url{http://arxiv.org/abs/1102.3151}

\bibitem{pradic25}
Pradic, C.: The equational theory of the {W}eihrauch lattice with (iterated)
  composition (2025), \url{https://arxiv.org/abs/2408.14999}

\bibitem{spivak2024reference}
Spivak, D.I.: A reference for categorical structures on $\mathbf{Poly}$ (2024)

\bibitem{streicherfibrations}
Streicher, T.: Fibered categories à la {J}ean {B}énabou (2023)

\bibitem{TVdP22}
Trotta, D., Valenti, M., de~Paiva, V.: Categorifying computable reducibilities.
  arXiv preprint arXiv:2208.08656  (2022)

\bibitem{UFTRING20}
Uftring, P.: The characterization of {W}eihrauch reducibility in systems
  containing $\mathrm{E-PA}^\omega + \mathrm{QF-AC}^{0,0}$. The Journal of
  Symbolic Logic  \textbf{86}(1),  224–261 (Oct 2020).
  \doi{10.1017/jsl.2020.53}, \url{http://dx.doi.org/10.1017/jsl.2020.53}

\bibitem{vanOosten}
Van~Oosten, J.: Realizability: an introduction to its categorical side,
  vol.~152. Elsevier (2008)

\bibitem{westrick2020}
Westrick, L.: A note on the diamond operator. Computability  \textbf{10}(2),
  107--110 (2021). \doi{10.3233/COM-200295}

\bibitem{Yoshimura2}
Yoshimura, K.: General treatment of non-standard realizabilites (2016),
  unpublished

\end{thebibliography}
\end{document}